# BK LYNCIS: THE OLDEST OLD NOVA?... AND A BELLWETHER FOR CATACLYSMIC-VARIABLE EVOLUTION

by

Joseph Patterson[1], Helena Uthas[1], Jonathan Kemp[1], Enrique de Miguel[2],

Thomas Krajci[3], Jerry Foote[4], Franz-Josef Hambsch[5], Tut Campbell[6],

George Roberts[7], David Cejudo[8], Shawn Dvorak[9], Tonny Vanmunster[10],

Robert Koff[11], David Skillman[12], David Harvey[13], Brian Martin[14], John Rock[15],

David Boyd[16], Arto Oksanen[17], Etienne Morelle[18], Joseph Ulowetz[19],

Anthony Kroes[20], Richard Sabo[21], and Lasse Jensen[22]

---

1  Columbia University, Department of Astronomy, 550 West 120th Street, New York, NY 10027; jop@astro.columbia.edu, helena@astro.columbia.edu, jonathan@astro.columbia.edu
2  CBA-Huelva, Observatorio del CIECEM, Matalascañas, 21076 Almonte, Huelva, Spain & Departamento de Física Aplicada, Universidad de Huelva, 21071 Huelva, Spain; edmiguel63@gmail.com
3  CBA New Mexico, PO Box 1351 Cloudcroft, NM 88317; tom_krajci@tularosa.net
4  CBA-Utah, 4175 East Red Cliffs Drive, Kanab, UT 84741; jfoote@scopecraft.com
5  CBA-Mol, Oude Bleken 12, B-2400 Mol, Belgium; hambsch@telenet.be
6  CBA-Arkansas, 7021 Whispering Pine, Harrison, AR 72601; jmontecamp@yahoo.com
7  CBA-Tennessee, 2007 Cedarmont Drive, Franklin, TN 37067; georgeroberts0804@att.net
8  CBA-Madrid, Camino de las Canteras 42, Buzon 5, La Pradera del Amor, El Berrueco, 28192 Madrid, Spain; davcejudo@gmail.com
9  CBA-Orlando, Rolling Hills Observatory, 1643 Nightfall Drive, Clermont, FL; sdvorak@rollinghillsobs.org
10 CBA-Belgium, Walhostraat 1A, B-3401 Landen, Belgium; tonny.vanmunster@gmail.com
11 CBA-Colorado, Antelope Hills Observatory, 980 Antelope Drive West, Bennett, CO 80102; bob@antelopehillsobservatory.org
12 CBA-East, 6-G Ridge Road, Greenbelt, MD, 20770; dskillman@comcast.net
13 CBA-West (deceased)
14 King's University College, Department of Physics, 9125 50th Street, Edmonton, AB T5H 2M1, Canada; brian.martin@kingsu.ca
15 CBA-Wilts, 2 Spa Close, Highworth, Swindon, Wilts SN6 7PJ, United Kingdom; john@highworthobs.fsnet.co.uk
16 CBA-Oxford, 5 Silver Lane, West Challow, Wantage, OX12 9TX, United Kingdom; davidboyd@orion.me.uk
17 CBA-Finland, Verkkoniementie 30, FI-40950 Muurame, Finland; arto.oksanen@jklsirius.fi
18 CBA-France, 9 Rue Vasco de Gama, 59553 Lauwin Planque, France; etmor@free.fr
19 CBA-Illinois, 855 Fair Lane, Northbrook, IL 60062; joe700a@gmail.com
20 CBA-Wisconsin, Cedar Drive Observatory, W2175 Cedar Drive, Pulaski, WI 54162; akroes@netnet.net
21 CBA-Montana, 2336 Trailcrest Drive, Bozeman, MT 59718; richard@theglobal.net
22 CBA-Denmark, Søndervej 38, DK-8350 Hundslund, Denmark

ABSTRACT

We summarize the results of a 20-year campaign to study the light curves of BK Lyncis, a nova-like star strangely located below the 2-3 hour orbital period gap in the family of cataclysmic variables. Two apparent "superhumps" dominate the nightly light curves – with periods 4.6% longer, and 3.0% shorter, than $P_{orb}$. The first appears to be associated with the star's brighter states ($V$~14), while the second appears to be present throughout and becomes very dominant in the low state ($V$~15.7). It's plausible that these arise, respectively, from a prograde apsidal precession and a retrograde nodal precession of the star's accretion disk.

Starting in the year 2005, the star's light curve became indistinguishable from that of a dwarf nova – in particular, that of the ER UMa subclass. No such clear transition has ever been observed in a cataclysmic variable. Reviewing all the star's oddities, we speculate: (a) BK Lyn is the remnant of the probable nova on 30 December 101, and (b) it has been fading ever since, but has taken ~2000 years for the accretion rate to drop sufficiently to permit dwarf-nova eruptions. If such behavior is common, it can explain other puzzles of CV evolution. *One:* why the ER UMa class even exists (because all members can be remnants of recent novae). *Two:* why ER UMa stars and short-period novalikes are rare (because their lifetimes, which are essentially cooling times, are short). *Three*: why short-period novae all decline to luminosity states far above their true quiescence (because they're just getting started in their postnova cooling). *Four*: why the orbital periods, accretion rates, and white-dwarf temperatures of short-period CVs are somewhat too large to arise purely from the effects of gravitational radiation (because the unexpectedly long interval of enhanced postnova brightness boosts the mean mass-transfer rate). And maybe even *five*: why very old, post-period-bounce CVs are hard to find (because the higher mass-loss rates have "burned them out"). These are substantial rewards in return for one investment of hypothesis: that the second parameter in CV evolution, besides $P_{orb}$, is **time since the last classical-nova eruption.**

## 1. INTRODUCTION

BK Lyncis was discovered in the Palomar-Green survey for objects with ultraviolet excess (Green et al. 1986), and was listed as PG0917+342 in the preliminary catalog of cataclysmic-variable stars in that survey (Green et al. 1982). A subsequent radial-velocity study confirmed the CV identification and revealed an orbital period of 107.97 minutes (Ringwald et al. 1996). Two years of time-series photometry revealed "superhumps" in the star's light curve – large-amplitude waves interpreted as resulting from apsidal precession of the accretion disk (Skillman & Patterson 1993, hereafter SP). These studies showed only small variability in the range $V$=14.5-14.7. Thus the star became well-established as a "novalike variable", a class which would be unremarkable, except for the star's short orbital period. Of the several hundred CVs known with orbital period below 2 hours, BK Lyn is the **only** novalike variable.

BK Lyn is also a good candidate as the "oldest old nova". Several studies of ancient Chinese records have suggested that a nova appeared very close to its position on 30 December 101 (Hsi 1958, Pskovskii 1972, Clark & Stephenson 1977), and Hertzog (1986) concluded that BK Lyn is the remnant of Nova Lyncis 101. This would certainly qualify as the oldest old nova – far exceeding the closest challenger, WY Sge = Nova Sagittae 1783 (Shara et al. 1985).

Such curiosities have kept BK Lyn on our observing lists for years. In this paper we summarize the results of many observational campaigns: spanning 20 years, and including ~400 nights and ~2200 hours of time-series photometry. Among the several rewards, detailed here, was the star's spectacular transformation into a *bona fide* dwarf nova in 2011-12. That transformation may provide a powerful clue to the long-term evolution of cataclysmic variables.

## 2. OBSERVATIONAL TECHNIQUES

Essentially all the data reported here comes from the Center for Backyard Astrophysics, a global network of telescopes cooperating in campaigns of time-series photometry of variable stars (CBA: Patterson 2006). Most of the observational techniques were discussed in the second paper of our series (SP), but the network expanded in later years to include ~20 telescopes, spread sufficiently over the Earth to give very long time series relatively untroubled by local weather and daily aliasing. Our typical telescope is a 35 cm reflector, equipped with a CCD camera and recording images every 60 s for many hours per night. Most of the data is unfiltered (white-light, or perhaps more correctly "pink", with an effective wavelength near 6000 A) differential photometry, although we always obtain some coverage in $V$ light to express results on a standard scale. Data from several telescopes are then spliced together to form a one-night light curve, with minimal gaps. We take advantage of overlaps in data to determine additive constants which put all our measurements on one instrumental scale (usually that of the most prolific or best-calibrated observer). These constants are usually in the range 0.01-0.05 mag, probably due to variations in transparency and camera sensitivity. Most telescopes use the same comparison star, although we also use data with other comparisons (requiring larger and more uncertain additive constants) if there is sufficient overlap. In this case we used GSC 2496-00893, which is 3.6 arcmin NE from BK Lyn. On 5 good nights in 2011, we measured that star to have *V*=13.897(8), *B-V*=0.533(10).

Research programs on faint stars with small telescopes often use white light, to enable high time resolution with good signal-to-noise. In the case of cataclysmic variables, it usually makes good astrophysical sense too, since the underlying sources of light are broad-band emitters (accretion disk, white dwarf). It is common practice to report magnitudes as "C" (or often "CV", though we will avoid this term for obvious reasons): the result of differential photometry in clear light, added to the comparison star's known *V* magnitude. This is also our practice. However, because the white-light passbands are typically ~4000 A wide, the effective wavelengths of the variable and comparison stars can easily be 500 A apart. Therefore, C/CV magnitudes are not *V* magnitudes. We nevertheless prefer the C/CV scale and use it here, because it is our natural measurement scale, and because it accurately expresses the true changes in light.

Since an instrumental scale is not fully reproducible, a standard *V* magnitude is more desirable for archival purposes. For "good" comparison stars (*B-V*<1.0), our *C* magnitudes transform to *V* magnitudes via

$$\Delta V = \Delta C + 0.37\, \Delta(B-V),$$

which implies $\Delta V = -0.20$ in this case, where the variable is assumed (and observed) to have *B-V* near 0.0. The latter assumption is pretty good for the great majority of cataclysmic variables accreting at a high rate – including BK Lyn.

Atmospheric extinction is significant for us, because the program stars are usually much bluer than comparison stars (although we avoid very red stars, which are the bane of all stellar photometry). We know from experience that this differential extinction amounts to ~0.06 mag/airmass for most CVs. Nevertheless, in the spirit of keeping human hands off the data as much as possible, we usually make *no correction for extinction*.

The summary observing log in the five new observing seasons (adding to the two seasons reported by SP) is given in Table 1. (A "night" denotes a time-series of good quality lasting at least 3 hours.)

## 3. SEASONAL LIGHT CURVES

During all years prior to 2011, our observations seemed consistent with the star's billing as a novalike variable, showing small excursions about a mean $V$=14.6. This is also consistent with the snapshot *Roboscope* data reported by Ringwald et al. (1996), which found the star always in the range 14.6-14.7 on 116 nights in 1994 and 1995. And it's consistent with the range (14.5-14.8) listed in the General Catalogue of Variable Stars (GCVS).

But in 2011, the star clearly had excursions to a fainter state (15.5), as well as a brighter state (13.5). This seasonal light curve is shown in the upper frame of Figure 1. In 2012, the variations were closely monitored over a long baseline, and revealed a pattern which bore all the earmarks of a dwarf-nova – in particular, a dwarf nova of the ER UMa class. This is shown in the middle frame of Figure 1. In fact, by coincidence we were simultaneously carrying out a long monitoring campaign on ER UMa itself, and that seasonal light curve was practically indistinguishable – even in small details – from that of BK Lyn in 2012. This is shown in the bottom frame of Figure 1. In both cases, apparent superoutbursts occurred every 45 days, followed by smaller excursions – possibly "normal" outbursts[23] – repeating every 4-5 days. Furthermore, as will be discussed in the next section, the two stars revealed a complex and identical morphology of periodic signals. It appears that BK Lyn, after many years of living as a novalike variable, became a dwarf nova by 2011.

## 4. NIGHTLY AND SPLICED LIGHT CURVES

Our main program was to study the time series for periodic signals. The first two seasons have been published (SP: 1992 and 1993). BK Lyn then stayed always near $V$=14.6, and showed waves with a period slightly exceeding $P_{orb}$: a "positive superhump". The average full (peak-to-trough) amplitude was ~0.07 mag, and the period excess was 4.6%, fairly typical for superhumps in CVs of comparable $P_{orb}$.

Later years showed a much stronger signal, which we illustrate in Figure 2, a 4-day segment obtained in 2012. This dominated all the later data (1999-2012), and was especially strong, approaching 0.6 mag full amplitude, when the star became faint. The period was 3.0% shorter than $P_{orb}$, which therefore signified a *negative* superhump (since the period excess is negative). In this section, we report on this and other signals in the various years' light curves.

In what follows, we sometimes adopt a date convention of truncated Julian dates ("JD" = true JD – 2,400,000), and a frequency convention of cycles $d^{-1}$ = c/d (for compactness, and as the natural unit of frequency in programs affected by daily aliasing). Since this paper is equally a story of classical and dwarf novae, we also need a convention for describing their shenanigans: classical novae have *eruptions*, whereas dwarf novae will be described as having "outbursts" or "maxima". Finally, we use the term *novalike* to describe noneruptive and nonmagnetic CVs whose spectrum, excitation, and $M_v$ are similar to the prototype, UX Ursae Majoris. (This is the most common use of the term, although some authors use it more expansively[24] – to describe any CV not known to be a dwarf nova, old nova, or magnetic.)

---

23 This term is meant to be merely descriptive, not indicative of the origin. These short outbursts are also sometimes called "reflares" or "echo outbursts"; their actual origin is still not securely understood.

### 4.1  1999 Campaign

The 1999 campaign spanned 26 days, with good instrumental magnitudes (internally calibrated on a delta-magnitude scale) and a dense segment during JD 51218-32. The star was always near $V \approx 14.6$. The power spectrum of this dense segment is shown in the upper frame of Figure 3. Strong signals occur at 0.400(9) and 13.739(9) c/d; these are sensibly phase-stable over the 26 days, and have full amplitudes – respectively – of 0.17 and 0.13 mag. After subtracting these two powerful signals, the time series shows weaker but significant features at 12.770(9) and 25.560(9) c/d. Shown in the middle frame of Figure 3, these latter signals are apparently a manifestation of the positive superhump, familiar from 1992/3. The strong signals are new – the first definite appearance of a negative superhump in our data. The latter are a fairly common phenomenon in CVs of high accretion rate[25]. The mean waveforms of both superhumps are shown in the lowest frame. The negative superhump is closely sinusoidal, while the positive superhump has a strong second harmonic.

Unfortunately, the negative and positive superhumps are separated by a frequency close to 1 c/d, a potentially cruel blow to astronomers on our planet. Still, the star never promised us a rose garden... and this long campaign with good alias rejection was able to separate the two signals.

At the risk of injecting some interpretation into matters of reportage (seldom wise), we will call the positive superhump "apsidal" and express the frequency as ω-A... and the negative superhump "nodal" and express the frequency as ω+N (where ω is the orbital frequency). This terminology adopts the common opinion that the accretion disk is eccentric and undergoes prograde apsidal precession at a rate A... and that it is also tilted away from the orbital plane, undergoing retrograde nodal precession at a rate N.  This would be a natural consequence of the secondary's perturbation if the disk were simply one orbiting particle (as with the Moon's orbit) – but is basically hypothetical for a structure as complex as an accretion disk.  This interpretation has become common (dating back to Barrett et al. 1988, Harvey et al. 1995, Patterson et al. 1997), and supported by theoretical work dating back to Whitehurst (1988) and Lubow (1991, 1992).  Good recent accounts of the theory have been given by Montgomery (2009, 2012) and Wood & Burke (2007).

---

24 For example, some authors use the term to describe any nonmagnetic stars not yet known to show dwarf-nova outbursts. This has merit for long-$P_{orb}$ stars, many of which have ~50-year records with no outbursts. But nonmagnetic short-$P_{orb}$ CVs will almost always show a dwarf-nova outburst, if you keep watching for ~5-20 years. "Novalike" thus usually proves to be the wrong classification for such stars.
25 Of a *sustained* high accretion rate. Practically all superhumps are born in states of high accretion. Dwarf novae teach us that a few days of high accretion suffice to hatch positive superhumps; but negative superhumps are mainly found in novalikes, suggesting that their growth time is probably much longer.

With that convention, the 1999 signals occur at frequencies N, $\omega$-A, $\omega$+N, 2$\omega$-2A, and 2$\omega$-A.  In fact, for all years of observation, the detected frequencies obey these simple rules:
   (1)  all apparitions of N are in the form +N, +2N, etc.;
   (2)  all apparitions of A are in the form -A, -2A, etc.;
   (3)  whenever a +N sideband appears, a strong low-frequency signal at N appears;
   (4)  whenever a -A sideband appears, a low-frequency signal at A never appears.
In fact, for virtually all our data on *all* superhumping CVs, these rules appear to be quite general – although data quality is sometimes poor for frequencies below 2 c/d (so upper limits for the power at N and A can be rather coarse).

This terminology will help our reportage in this complex story of periodic signals.  A primer on periodic-signal and superhump zoology in cataclysmic variables can be found in Appendix A of Patterson et al. (2002).

### 4.2  2002

The 2002 campaign spanned 58 days, with a dense segment during JD 52264-318.  No substantial difference from the 1999 results was found.  The powerful signals occurred again at 0.393(2) and 13.738(2) c/d.  After subtracting these, weaker signals appeared at 25.521(2), 27.475(2), 12.760(2), and 13.344(2) c/d, in order of decreasing power.  With the convention described above, these are detections of the N, $\omega$+N, 2($\omega$-A), 2($\omega$+N), $\omega$-A, and $\omega$ signals – in order of decreasing power.  The basic underlying clocks are then measured to be $\omega$=13.344(2), N=0.395(2), and A=0.584(2) c/d.  If our interpretation is correct, the orbital frequency $\omega$ should be absolutely stable, while N and A could vary slightly, since they are characteristic of the accretion disk – a much more loosely organized structure than a binary orbit.  Some higher harmonics and sidebands are also seen; these are detailed in Table 2.

### 4.3  2005

The 2005 campaign spanned 20 days, when the star appeared to stay close to *V* = 14.8.  The runs were sufficiently short that no reliable analysis was feasible for signals of very low frequency.  But the power spectrum showed the usual superhump signals at 13.740(4) and 12.751(4) c/d and their second harmonics, with full amplitudes of 0.140 and 0.024 mag, respectively.  The superhump waveforms were also quite similar to those of 1999 and 2002.

While writing this paper, we became aware (thanks to Patrick Wils) of the star's brightness history in the Catalina Real-Time Transient Survey. With coverage starting in 2005, that history shows rapid variability in the range *V*=14 to 16. The data (~20 snapshot magnitudes per year) are too sparse to reveal a period or to certify "dwarf nova", but establishes that the pattern of up-and-down variations – obvious in later years and clearly arising from dwarf-nova activity – began as early as 2005. In retrospect, the 2005 CBA coverage was apparently concentrated near the "plateau" phase of a superoutburst, which concealed the full range of variability.

### 4.4  2011

The star's 2011 light curve covered 324 hours and 58 nights. The seasonal light curve, seen in Figure 1, shows the star to have had significantly different brightness states – both fainter and brighter than anything seen by us in previous years. This complicated the analysis of periodic signals, and we defer that analysis to a later paper. We resolved to do a more thorough job in the 2012 campaign.

### 4.5  2012

And so we did. Table 1 shows details of the coverage (1108 hours), and Figures 1 and 2 show the seasonal and several-day light curves. Recurrent high and low states are now obvious. Interpreted with dwarf-nova terminology, the seasonal light curve suggests superoutbursts every 45±3 days, and normal outbursts every 5.3±0.6 days. Such a dichotomy is a hallmark of short-$P_{orb}$ dwarf novae, and the frantic pace (45 and 5 days) is a distinctive hallmark of the ER UMa subclass. Robertson et al. (1995) gives a good observational account of this class; and Osaki (1995, 1996) gives a lucid explanation of the high frequency of outburst: exceptionally high $\dot{M}$.

The large fluctuations in brightness created problems in periodicity analysis, but the basic patterns are easily summarized:

1. A positive (apsidal) superhump grows suddenly to very large amplitude (0.3 mag) at the beginning of each superoutburst, and decays slowly as the outburst does, over ~10 days. Its waveform initially shows the nearly universal fast-rise-slow-decline pattern, but then mutates to a double-humped shape after a few days.

2. A negative superhump is present all the time, with an amplitude which is practically constant in intensity units (~0.5 mag at *V*=16, 0.07 mag at *V*=14). This phase and amplitude (in intensity units) appear to pay no attention to any outbursts which may be occurring. And it always dominates when the star is faint; the pattern obvious in Figure 2 repeats almost exactly every time the star shuffles between bright and faint states.[26]

In the following, we show the basis for these summary points. The first is a characteristic of essentially all short-period dwarf novae. The second is closely reproduced by several ER UMa stars: most prominently, V503 Cygni (Harvey et al. 1995) and ER UMa itself (Ohshima et al. 2012, de Miguel et al. 2012). It has never been conclusively observed in any other type of dwarf nova.

Our global telescope network can distinguish between these two signals, despite the unlucky separation in frequency (0.98-1.01 c/day). However, not every subset of the data *independently* distinguishes between them. So we have generally parsed the time series into three categories: superoutburst, quiescence, and normal outburst – and studied each separately.

### 4.5.1 Periodic Signals in Superoutburst

During each of the three well-observed superoutbursts, we analyzed the data in the manner we usually apply to erupting dwarf novae: by subtracting the mean light from each daily time series ("zeroing"), splicing to form a ~10-day light curve, and then calculating the power spectrum from a discrete Fourier transform. A seven-day light curve covering the first superoutburst is shown in the upper frame of Figure 4, and the power spectrum is shown in the lower frame. Both $\omega$-A and $\omega$+N appear prominently, plus some higher harmonics and linear combinations. Study of the light curves showed an obvious pattern: the $\omega$-A signal became suddenly very strong at the peak of superoutburst, and then decayed away over a few days.

Because of the unlucky coincidence in frequency, we could not separate the two superhump signals on each *individual* night, and therefore could not confidently distinguish the amplitudes on every single night. Nevertheless, the pattern described above was obvious in each superoutburst.

### 4.5.2 Periodic Signals in Quiescence and Normal Outburst

Away from superoutburst, the dominant signal was always the negative superhump, and on each night we measured the mean amplitude and phase (the time of maximum light, or "pulse arrival time"). Figure 5 shows the dependence of amplitude on the star's brightness, in agreement with the segment shown in Figure 2 and the description given above (with outlier high amplitudes near maximum light at *V*=14; these arise from temporary contamination by the apsidal superhump). The times of maximum light are given in Table 3, and the wandering of the phase – relative to the mean period – is tracked by the O-C diagram of Figure 6. Apparently the phase wanders by ~0.1 cycle on a timescale of weeks – but not enough to lose cycle count across the outbursts. This enables a very accurate measure of the period: 0.072846(5) d, or ω+N = 13.7276(9) c/d. The period is 0.09(2)% longer than it was in 2002, the other season of accurate measurement. This instability in period gave us confidence that the proper description of the signal is "negative superhump" rather than "spin period of the white dwarf" (which would be much more stable).

---

26 The Kepler light curves of V344 Lyr (Wood et al. 2011, especially their Figures 10 and 11) show this general pattern beautifully.

We searched for a signal near 0.4 c/d, the required location of N. For such low frequencies, we must use the actual magnitudes (without a subtraction of the mean), and therefore the normal outbursts contaminated the power spectra with enormous noise at low frequencies. This prevented any highly accurate measure of the low-frequency signal; but during several stretches of relatively constant light, there were obviously significant peaks near 0.39-0.40 c/d.

## 5. SUMMARY OF PERIODS

Table 2 contains a summary of the periodic signals detected (in frequency units, to enable a clearer parsing among superhumps/harmonics). Each identified signal is fairly stable in frequency, but the years of long-baseline coverage show that differences are measurable: the apsidal superhump apparently increased by ~0.2% in frequency between 1992-4 and all subsequent years, and the nodal superhump apparently decreased in frequency by ~0.1% between 2002 and 2012.

Superhumps presumably originate from perturbations of the disk by the secondary, and in particular of the disk's outer regions, since the perturbations are much stronger there. And when the star declines in light, it probably means that $\dot{M}$ and the outer disk radius are declining too. A decline in $R_{disk}$ weakens the perturbation and therefore should move both the apsidal and nodal superhump frequencies closer to $\omega_{orb}$. This would increase $\omega_A$ and decrease $\omega_N$, as observed. Thus it's possible that exact superhump frequencies – measured over a baseline sufficiently long to smooth over accretion-disk "weather" – are a good proxy for $\dot{M}$, and that the entries in Table 2 signify a slight decrease of $\dot{M}$ over the 20 years of observation.[27]

## 6. BK LYN IN THE DWARF-NOVA FAMILY

The seasonal light curve and the periodic signals (ω, N, A, and their children) leave no doubt that BK Lyn was a fully credentialed dwarf nova in 2012. Let's review some salient points about short-$P_{orb}$ dwarf novae.

---

27 A similar trend – anticorrelated changes in $\omega_A$ and $\omega_N$ – was also noted in V603 Aql (see §8 and Figure 8 of Patterson et al. 1997). That provides some comfort to the disk-precession theory. But for both stars, night-to-night brightness variations were too great to permit directly testing the expected correlation with $\dot{M}$.

Nearly all well-studied dwarf novae of short $P_{orb}$ show a long/short dichotomy in their outbursts. During all long ("super") outbursts, a strong apsidal superhump (ω-A) is quickly born, and decays after a few weeks – roughly, but not exactly, when the outburst does. Short eruptions never hatch such signals. Stars like this are called "SU UMa-type" dwarf novae. There are ~200 such stars, and therefore a few hundred outbursts sufficiently studied to test the universality of these statements. Only a few stars (<5%) have not yet clearly professed allegiance to these patterns. Warner (1985) and Patterson et al. (2005, hereafter P05) present reviews of these stars, and Kato et al. (2010, 2011) present large collections of data.

Present-day nomenclature also ordains subclasses. The superoutbursts of so-called WZ Sge stars happen very rarely (P>10 years), while those of ER UMa stars happen very frequently (P<120 days). Stars in between are just plain old SU UMas. In our opinion, these subclasses do not reflect any essential difference in physics, but merely accretion rate. Osaki (1996, see his Figure 2) shows simply and lucidly why outburst recurrence rate should vary smoothly with accretion rate; Patterson (2011, hereafter P11) shows that observations bear this out, and in particular that recurrence period scales with <$M_v$>, which is a proxy for $\dot{M}^{-1}$ (Figures 7 and 11 of that paper).

Superhumps are a great distinguishing feature of all such stars, and are sometimes taken to be a *defining* feature of the SU UMa class. Our enthusiasm doesn't go quite that far, however. It is more general, and more interesting, to say that (apsidal) superhumps inevitably result when stars of sufficiently short $P_{orb}$, and containing accretion disks, achieve sufficiently high $\dot{M}$ for a sufficiently long time. It's then up to observers to determine what constitutes "sufficient". These sufficiency conditions were estimated by P05 as follows: $P_{orb}$<3.5 hours, $\dot{M}$ ≈ 3x10$^{-9}$ $M_o$/yr, t ~ a few days. For short-period ($P_{orb}$<2.4 hr) dwarf novae, these conditions appear to be always satisfied in superoutburst, never in normal outburst, and never in quiescence. It's amazing how faithfully the stars follow these rules!

In its long/short dichotomy of outburst, superoutburst interval, $P_{orb}$, <$M_v$>, and rise/fall pattern of the apsidal superhump, BK Lyn is a fully credentialed member of the ER UMa subclass. This is underlined further by the negative superhumps, because among dwarf novae, these appear to be a property special to ER UMas (found in at least 6 out of 10, compared to ~2 out of ~300 for other dwarf novae).[28]

## 7. BK LYN AS A CLASSICAL NOVA

The "guest star" of 101 A.D. has been previously discussed by Hsi (1958),

---

[28] The precise numbers depend on exactly how ER UMa membership is defined (we have adopted $T_{super}$<120 d), and how clear is the evidence for negative superhumps.

Ho (1962), Clark & Stephenson (1978), and especially Hertzog (1986). Unlike most guest stars, it is at high galactic latitude (44°), and can be placed with unusual precision in the sky, since it is described as very close to a star ("the fourth star of Hsien-Yuan") which all students of ancient records take to be Alpha Lyncis. Hertzog argues that a compelling case can be made for BK Lyn, a scant 29 arcminutes away from α Lyn. The modern-day BK Lyn is an extremely unusual star – the only short-period novalike among the ~1000 known CVs, and now the only novalike to have clearly morphed into a dwarf nova. Accurate positional coincidence with a very unusual object constitutes some evidence of physical association.

The Chinese records describe the star as "small", which, in the context of other brightness reports in those records, is taken to mean a magnitude near zero (Hertzog 1986). Superoutbursts of the modern-day dwarf nova reach $V$=13.9 (allowing for the difference between $V$ and $C$ magnitudes), and superoutbursts of dwarf novae are pretty good "standard candles" with $M_v$=+4.5 at maximum light (Figure 1 of P11). Allowing an additional 0.2 mag for absorption on this line of sight, and a reward of ~0.3 mag for the likely low binary inclination[29], we estimate a distance of ~800 pc. If the guest star is actually BK Lyn, then it apparently rose to $M_v$ = -9.7 (0 minus 0.2, with a distance modulus $m-M$ = 9.5). That's about right for a fast classical nova, and a cataclysmic variable is exactly what's needed as a classical-nova progenitor. Finally, we note that BK Lyn has just executed the move which theorists have long predicted must occur for very old novae: it has settled back into a dwarf-nova state. These arguments from physics and brightness strengthen the argument from positional agreement, and we conclude, following Hertzog, that BK Lyn is likely to be the remnant of Nova Lyn 101.

## 8. THE DECLINE OF CLASSICAL NOVAE

Identification of BK Lyn with an ancient nova could give some powerful clues to the evolution of novae. One concerns the question: how long do novae stay bright?

In a classic study of old novae, mainly from the 20th century, Robinson (1975) found that novae usually have the same brightness before and after eruption. But since most known novae arise in stars of long $P_{orb}$, this finding only applies to that class. Expressed in terms of accretion rate, this implies that (long-$P_{orb}$) postnovae fade to ~$10^{-8}$ Mo/yr within 10-20 years after eruption... and then, within the limits of data on photographic surveys, seem to have the same brightness, and therefore accretion rate, a few years before the eruption. The simplest interpretation is that these stars are similarly bright throughout the long interval between eruptions. We strongly suspect that this is true – because the several hundred presently known nonmagnetic long-$P_{orb}$ CVs, which are presumably the ancestors of most future novae, are all similarly bright (see Figure 7 of P84; the several low points on this figure at long $P_{orb}$ have all acquired excuses: magnetism or subsequently observed outbursts). Among long-$P_{orb}$ stars, prenovae, postnovae, and nova suspects all look about the same (aside from fireworks associated with the eruption itself: ejected gas and dust shells, supersoft X-rays, etc.). This is why *novalike* has become a common and useful term: because in the long-$P_{orb}$ regime, *the spectra and light curves[30] of stars not known to be old novae (or dwarf novae) are basically indistinguishable from those of the old novae*.

---

29 The standard-candle constraint assumes the accretion disk to be inclined at an average angle, namely 57°. We're unable to measure the binary inclination, but some constraints are available: BK Lyn is noneclipsing, with moderately narrow lines and a very weak (~0.01 mag) orbital modulation. From this we estimate $i$ in the range 20-60° – and adopt $i$=45°, which implies an $M_v$ correction of 0.3 mag (and thus $M_v$ = +4.2 at maximum light).

So: apart from the 10-20 year aftermath of the eruption, long-$P_{orb}$ novae quickly settle to a long quiescence near $M_v = +5$, $\dot{M} = 10^{-8}$ $M_o$/yr (see Figures 7 and 12, and Table 5, of P84).

The situation for short-$P_{orb}$ stars should be, and is, radically different. Why? *Because these stars are not naturally entitled to accrete at $10^{-8}$ $M_o$/yr*. Their only known driver of mass transfer is gravitational radiation (GR), which only provides $10^{-10}$ $M_o$/yr. Therefore they have to wait ~100x longer[31] to pile up enough matter to fuel another eruption, and that may be plenty of time to cool sufficiently to join their natural compadres at short $P_{orb}$ – the garden-variety dwarf novae, with a quiescent $M_v$ near +9.5 and brief outbursts every 150-2000 days.

In this scenario, BK Lyn is the product of a recent nova eruption – probably, though not necessarily, the event of 30 December 101. With a mere 2000 years of cooling, the white dwarf is still fairly hot, and the secondary star sufficiently agitated to transfer matter at an unnaturally high rate – near $10^{-9}$ $M_o$/yr. This explains the high temperature component seen in the ultraviolet spectrum (35000 K, Zellem et al. 2009), and the star's intrinsic brightness (at 800 pc, the light curve implies $< M_v> =+5.7$ in 2012). Somewhere between 2002 and 2005, the star faded sufficiently to allow dwarf-nova outbursts to occur.

---

30 But not quite the absolute magnitude. The old novae average +4.1, whereas other long-$P_{orb}$ stars average about +5 (P84). This is consistent with the scenario peddled below, since the old novae erupted more recently.

31 And probably even longer. Townsley & Bildsten (2004) argue that the ignition mass should rise sharply for low accretion rates, causing that factor to be closer to 1000x (see their Figure 8). As they remark in their follow-up paper (Townsley & Bildsten 2005), both WD physics and nova statistics basically *require* a fairly sharp dependence on accretion rate. This is an important caveat. Likewise, accretion rates on either side of the period gap vary significantly with period (Figure 7 of P84). But in framing an argument, there is virtue in round numbers and stationary targets. For simplicity we adopt them in this paper: $10^{-8}$ $M_o$/yr above the period gap, $10^{-10}$ $M_o$/yr below, and an ignition mass of $10^{-4}$ $M_o$.

This appears to set an important and previously unknown timescale: ~2000 years to resume life as a dwarf nova, viz. of the ER UMa persuasion. But ER UMa stars are themselves quite rare; P11 estimated that they comprise only 1-2% of the population of short-period dwarf novae. If the ER UMa stage lasts ~10000 years, the numbers work out about right: 2000 for the BK Lyn era, 10000 for the ER UMa era, and 1,000,000 for dwarf-nova normalcy. That satisfies the relative space-density constraints (P11 and P84), and allows the correct amount of time for binaries accreting at $10^{-10}$ Mo/yr to accumulate the $10^{-4}$ Mo needed to trigger a nova eruption.

Shouldn't we then expect most short-period novae to be much fainter prior to eruption, contrary to Robinson's study? Yes – but that study concerned mainly long-period novae. In an important study of (mainly) archival photographic magnitudes of short-period novae, Collazzi et al. (2009) and Schaefer & Collazzi (2010, hereafter SC) found very different behavior: they were all *much fainter in the several decades before eruption*. That evidence is consistent with our account of very slow relaxation following a classical nova in a short-period CV.

SC invoked magnetism as a proposed explanation of this dichotomy, citing evidence that light curves of all the short-period novae show periodic signals at a non-orbital frequency (CVs certified as magnetic nearly always show a photometric signal at the white dwarf's spin frequency). But we have carried out long photometric campaigns on three of the five stars in question (V1974 Cyg, CP Pup, and RW UMi); and to our eye, none show that cited evidence[32]. They each show periodic signals at a non-orbital frequency, but the signals' low phase stability is characteristic of a clock mechanism seated in the accretion disk ("superhumps", or something closely related), rather than in WD rotation. If BK Lyn is actually the remnant of a recent nova, then add that to the score: superhumps 4, spin 0.

Two other post-novae figure in the SC hypothesis of magnetism at short $P_{orb}$: GQ Mus and T Pyx. For T Pyx, this was based in a possible 2.6 hour photometric signal detected in a 1996 campaign (Patterson et al. 1998). But in ~1000 hours of photometry in later years, including each observing season, we have never seen this signal re-appear; without confirmation, its evidentiary value must be reckoned as weak. For GQ Mus the original case for magnetism rested mainly on the strong soft X-ray emission (Diaz & Steiner 1994). But the soft X-rays turned off after

---

32 This is reasonably shown by previously published studies of CP Pup (Patterson & Warner 1998) and V1974 Cyg [Skillman et al. (1997), Olech et al. (2001)]. In addition, our recent unpublished and long campaigns on all three stars fail to reveal any evidence of a *stable* non-orbital frequency.

~10 years (Shanley et al. 1995, Orio et al. 2001), and recent work has shown that a long-lasting soft X-ray phase is common in novae, presumably because the WD manages to stay hot for a few years after the main eruption. In other words, this is a standard feature of an eruption, rather than a sign of accretion energy from a magnetically channeled stream. Also, no actual evidence for magnetism has been found: no polarization, no second period (just a powerful and stable 85 minute signal believed to signify $P_{orb}$), and no period change which could suggest an "asynchronous polar". Thus, evidence for magnetism in T Pyx and GQ Mus, always weak, has nearly evaporated.

In our judgment, the other three stars in the SC gallery of magnetics never showed evidence for magnetism in the first place. So we think the score is: superhumps 4, orbit-only 2, magnetism 0 (unsubstantiated, not "certifiably nonmagnetic").[33]

We advocate a simpler explanation for the dichotomy found by SC – arising simply from the main property of short-period CVs: low $\dot{M}$. This is illustrated by Figure 7, which purports to be a universal roadmap for the late decline[34] of fast classical novae. The points show the observed AAVSO visual light curves of two fast novae: V603 Aql (1918) and V1974 Cyg (1992), averaging over many points (dozens to hundreds) at selected intervals of dense coverage. To render these on an absolute magnitude scale, we adopt distance estimates of 380 pc and 1800 pc respectively (McLaughlin 1960, Chochol et al. 1997, Hachisu & Kato 2012). These are excellent stars for comparison, for many reasons:
(1) they're very well observed;
(2) they show smooth declines;
(3) their distances are pretty well determined by the "expansion parallax" of their nova shells;
(4) they exemplify long- and short-period stars (3.32 and 1.95 hrs respectively); and
(5) their postnova light curves show both positive and negative[35] superhumps, sometimes separate and sometimes simultaneous; this fine detail is shared with BK Lyn and the ER UMa stars, and shows that the accretion disk dominates the light.

---

33 This just refers to the short-period members of the SC gallery. The three long-period members – V1500 Cyg, V4633 Sgr, and V723 Cas – are a different story. The first is certifiably magnetic (from polarization evidence), the second is likely magnetic (as a likely asynchronous polar), and information concerning the third is still lacking. Our remarks are limited to the short-period stars, who, by virtue of being denied access to magnetic braking, will face a very different evolution – once their glory days as freshly erupted novae have passed.
34 For an excellent and very detailed study of the early decline, see Hachisu & Kato (2010).
35 The negative superhumps in V1974 Cyg are not yet in the public record, but are clearly evident in CBA coverage during 1998 (Emir et al. 2013).

Figure 7 shows that the two novae track each other pretty well for 1-2 years, and then seem to depart. The long-$P_{orb}$ curve needs to asymptote to $M_v \sim +5$, and it needs to do so fairly fast, since the star is due for another nova eruption in 10000 years. (This is because all nonmagnetic long-$P_{orb}$ CVs with disks accrete at fairly high rates.) The short-$P_{orb}$ curve must eventually descend much further; the 20-year record of V1974 Cyg suggests the beginning of that trend, and the point at "quiescence" (the $V>21$ limit inferred from the star's invisibility on the Palomar Sky Survey: Collazzi et al. 2009) must eventually be reached – before the star erupts again in 1,000,000 years. This long interval, from 100 to 100,000 years after eruption, is essentially *terra incognita*, and we suggest here that BK Lyn and its ER UMa relatives point the way to understanding it. Since they are all dwarf novae, we know distances, and therefore absolute magnitudes, pretty well; these are discussed by P11, and we list them here in Table 4. BK Lyn has $<M_v> = 5.7$, and the above discussion places it at 2000 years. That should be the beginning of the ER UMa era (of postnova cooling). The ER UMa stars average $<M_v> = 7.1$; their lifetime must be short, but probably at least 5x that of the BK Lyn stage, judging from the considerably greater abundance of ER UMas in the CV population. In Figure 7 we place this class at 15000 years.[36].

Some readers, seduced by the fast decline of a nova's initial fireworks, may find this decline to be absurdly slow. But actually, the decline suggested by Figure 7 is

$$dM_v/d(\log t) = 1.0,$$

identical to the value obtained by Duerbeck (1992) from studying the first 30-100 years of decline ("historical") in many novae, and slightly faster than the theoretical decline rate deduced by Duerbeck from the work of Smak (1989). We suggest here that this is a nova's natural and eternal decline rate. But most novae are of long $P_{orb}$; a declining nova reaches $M_v = +4$ pretty fast (30 years), and there it joins hundreds of long-period CVs not known to be associated with novae. This has led to a common perception that nova eruptions are "finished" after ~30 years.

Figure 7 suggests that the eruption's effect in short-period CVs, free from the masking presence of strong magnetic braking, goes on for many millennia. Probably *forever*. The evolution line in Figure 7 shows that CVs spend most of the inter-eruption cycle near true quiescence, but don't necessarily emit most of their light in that state. Diluted over the full million-year interval, the light from year 10 to year 100000 averages $M_v$ = +9.9 – about the same as the star naturally possesses at true quiescence (near $M_v$ = +10; see Figure 7 of P84, Figure 5 of P11). Since the dominant part of that extra light is emitted many millennia after the classical nova event, and since the stars characteristically show periodic signals which betray a disk origin, this light is very likely to be accretion-powered. Equal light implies roughly equal accretion, so the mean accretion rate, averaged over the million years, is about twice that of true quiescence. This has deep implications for evolution, which we will explore in §10 and 11.

---

36 These estimated lifetimes are essentially based on the numbers of known short-period CVs in these classes (BK/ER/SU, which occur in the P11 census with a ratio 1:10:300). Discovery of BK Lyn/ER UMa stars is certainly hampered by their small outburst amplitudes, but helped by their relatively high luminosities. In the absence of wisdom on how to correct for such selection effects, we just use the raw P11 counts.

## 9. THE TRANSITION TO DWARF NOVA: A SINGULAR EVENT?

Does the observed 2002-5 transition to a dwarf-nova state represent a singular event in BK Lyn's postnova evolution? Well, maybe. Our 20-year observation span, and the 3-year window for this transition, are short – but not ridiculously short compared to the putative 2000-year wait. Also, we obtain similar photometric coverage of many old novae and novalikes (at least 50), and we have not observed such a transition in any other star. So even as a singular event, not to be repeated until the next nova cycle, it does not seem wildly improbable.

But there is no need to hypothesize a singular event. Secular decline in $\dot{M}$ over a few thousand years could easily be punctuated by small fluctuations about the temporary mean; indeed, many cataclysmic variables show small luminosity variations on timescales of decades (Warner 1988, Richman et al.1994). It's plausible that such fluctuations could now be swinging the disk between states of stable and unstable accretion. Actually, that is our current understanding of Z Cam stars: dwarf novae near that threshold accretion rate, with their disks fluctuating irregularly between stable and unstable. If a transition back to steady light occurs, BK Lyn would be considered the first short-period Z Cam star.

If this latter version were correct, then we might well see other ER UMa stars – especially RZ LMi with its whirlwind 20-day superoutburst cycle – mutating temporarily into novalike variables, pausing slightly in their inevitable decline towards a long and simple life as a garden-variety dwarf nova.

## 10. RELATIVES IN THE CV ZOO

This hypothetical story of BK Lyn's rise and fall, from yesterday's classical nova to tomorrow's ordinary dwarf nova, needs to give a coherent account of other specimens in the CV zoo. We have done this to some extent in Sections 6 and 8 above. With an advance apology for trafficking heavily in the arcana of individual stars, we now do so in more detail.

## 10.1 Other Nova → Dwarf Nova Transitions

Is this the first classical nova found to evolve into a dwarf nova? Probably not, on several grounds. One: the event of 30 December 101 is not known with certainty to be a classical nova. Two: BK Lyn is not yet known with certainty to be its remnant. And three: there are other postnovae which have been described as showing dwarf-nova outbursts, especially GK Per (Nova 1901: Cannizzo & Kenyon 1986, Nogami et al. 2002) and V446 Her (Nova 1960: Honeycutt, Robertson, & Kafka 2002; Thorstensen & Taylor 2000; Schreiber et al. 2000).

Concerning the first two points, reaching certainty is difficult 2000 years later; but the arguments of Hertzog (1986) appear strong, especially the excellent positional agreement. The uniquely high accretion rate and WD temperature add to those arguments; in that $P_{orb}$ regime, *the only other stars that bright and hot are known nova remnants*. And the observed mutation to a dwarf-nova state adds an exclamation point. As for V446 Her, its credentials as a dwarf nova are questionable – consisting mainly of an historical light curve showing up-and-down excursions, but with no other known properties specific to dwarf novae.[37]

GK Per is an interesting case. It now shows large outbursts every ~900 days, which do look like dwarf-nova outbursts; and variable-star archives (AAVSO, AFOEV) suggest that they started around the year 1960-1970. This appears to grossly violate the "2000 years to resume dwarf-nova activity" rule, and to mildly violate the lesson from other old novae (since no other modern nova has apparently done this). But the proposed 2000-year rule only applies to short-period novae, where the threshold is at $10^{-9}$ Mo/year, compared to $8 \times 10^{-9}$ Mo/year for an ordinary nova with $P_{orb}$ = 6 hr[38]. If all classical novae decline at one rate, then long-period stars reach their threshold sooner, enabling the stars to reach their final resting place (novalike or dwarf nova) sooner. This is well illustrated by Schreiber & Gaensicke (2001), especially their Figure 4. GK Per's very long $P_{orb}$ of 48 hr makes it a good candidate to reach its threshold rather fast.

---

37 Unfortunately, dwarf novae of long $P_{orb}$ sometimes lack the distinctive classification clues of their short-period cousins: superoutbursts and superhumps, each with very clear morphologies and time-dependences. At long $P_{orb}$, the term *dwarf nova* often means something less definite: "roughly cyclic variations in brightness". V446 Her satisfies this important criterion. But dwarf-nova outbursts generally show other properties: rapid rise in light from a fairly flat quiescence, sudden appearance of absorption lines, disappearance of flickering. None of this information is yet available for V446 Her.

38 These estimates are based on the full-disk calculations of Osaki [1996, his Eq. (4)]. However, GK Per itself is an awkward comparison star, because its WD is sufficiently magnetic to disrupt the inner disk, where most of the energy is released. So the quantitative argument is slightly murky, but the point is not: long-period stars don't have to wait as long.

Another star relevant to these matters is the dwarf nova Z Cam. Narrow-band and ultraviolet imaging has revealed filamentary emission near Z Cam, and these are interpreted as the remnants of a classical-nova shell ejected at least 1300 years ago (Shara et al. 2007, 2012). The evidence looks pretty good.[39]. But as a long-$P_{orb}$ star, Z Cam's subsidence to a dwarf-nova state in ~1000 years brings no special surprise for evolution theory (although it certainly brings delight!). On the contrary: stars with natural machines for powering a steady $10^{-8}$ Mo/yr ("magnetic braking") had better reach their quiescent states pretty fast, because they'll be erupting again in ~10000 years. The decay time can't be as short as 100 years, or we would know plenty of dwarf novae among historical novae (the actual number is 0, 1, or 2); and it can't be as long as 10000, or there would be practically no long-period dwarf novae (which are very numerous). One thousand seems about right.

These numbers would be vastly different for short-$P_{orb}$ stars, which naturally accrete at rates 100x lower. Our interest lies almost entirely in that wing of the CV zoo, since that's where BK Lyn's (pre-2005) properties stand out as unique.

10.2  Other Short-period Novae, Especially T Pyx

Although BK Lyn was the only short-period star classified as "novalike", this is somewhat of a technicality, since there are a few stars which would certainly receive that classification, except for their known classical-nova eruption. These are CP Pup (1942), RW UMi (1956), GQ Mus (1983), V1974 Cyg (1992), and T Pyx (6 eruptions). The first four have all remained far above their pre-eruption brightness (Collazzi et al. 2009, SC) – consistent with the simple idea, presented here, that fading to quiescence requires at least 10000 years (and probably longer, based on Figure 7 and the arguments given above).

But T Pyx is hugely different from all the others – since it erupts every 25 years, and shines at $M_v \approx +1$ at its (extreme) version of "quiescence". Can this be consistent with the story peddled here? What makes T Pyx unique?

One possibility is WD mass. The theoretical models of Yaron et al. (2005, their Table 2) show the great sensitivity of the underlying thermonuclear instability to mass. At M=1.25 Mo, the WD erupts after accreting only $2 \times 10^{-6}$ Mo – for a wide range of accretion rates, including the very high rates (~$10^{-7}$ Mo/yr) that must apply to T Pyx in order to account for its great quiescent luminosity. Then the star erupts every 20 years. The WD would remain very luminous and hot, and the secondary could be very strongly irradiated, enabling it to sustain a high rate of mass loss indefinitely (Knigge et al. 2000). More generally, the decline curves of Figure 7 imply a high accretion rate for decades or centuries after eruption – and therefore also contain the possibility, if the rate is high enough, that the WD may erupt prematurely. That short-circuits the decline and traps the star in endless rapid eruptions. Such stars should be very rare – because high WD masses are, and because suicidal rapid burnout quickly removes the stars from the night sky.

---

[39] There is even a suggested association with a particular guest star seen in 77 B.C. (Johansson 2007), although the large positional uncertainty (hundreds of square degrees) makes that difficult to assess.

This account of T Pyx seems plausible, except that it predicts a very low ejected mass, whereas observations suggest that ~$10^{-4}$ Mo was ejected in the 2011 eruption (Nelson et al. 2012, Schaefer et al. 2012, Patterson et al. 2012). T Pyx continues to defy explanation.

## 11. POSTNOVA LIGHT AND CATACLYSMIC-VARIABLE EVOLUTION

Actually, the estimate for postnova light in § 8 probably underestimates its importance. Actual stars will fade to quiescence on a cooling timescale, not "immediately"; the postnova phase may well last to some degree past year 100000. And second, ultraviolet fluxes show a "white dwarf" component of 35000 K in BK Lyn (Zellem et al. 2009), and temperatures in other ER UMas (Table 4) distinctly higher than the ~15000 K typical of an average short-period CV (Townsley & Gaensicke 2009). Whether this is truly the WD or just a very healthy ultraviolet disk flux, it does suggest that the ultraviolet components of the ER UMa stars produce a greater bolometric correction. Both effects boost the importance of postnova light, compared to true quiescence. So we would describe the mean accretion rate[40] as "at least twice" that of true quiescence.

### 11.1 Consequences for CV Evolution

The foregoing calculation is rough, depending on some hard-to-estimate numbers (duration of the postnova phases, total time to the next eruption, absolute magnitudes). For the issue of assessing the importance of postnova brightness in the overall energy budget, the most critical question is whether the ER UMa class (including its shiny new member, BK Lyn) is actually an evolutionary phase in the nova cycle. We think the answer is *yes*, based on BK Lyn's transformation and its likely association with a probable ancient nova – and also based on the lack of alternative excuses (magnetism, odd WD mass[41], etc.) for the high luminosity of ER UMas. Next most critical is the question, how long does it last? Our answer is based on the P11 estimate of space density relative to the total population of short-period CVs (1-2%). Selection effects in discovery, and some vagueness in the ER UMa certification (there are borderline members, and no indisputable criteria for membership) certainly muddy the waters on this point. But unless all our interpretations are false, postnovae ought to behave qualitatively like Figure 7. And by our arithmetic, it's hard to draw a curve through the points without suspecting that the time-averaged light radiated during the postnova phase might equal or exceed the quiescent light of short-period CVs.

---

40 We prefer $M_{bol}$ because it is more directly linked to accretion rate. However, it incurs the additional uncertainty of whether to separately parse the flux into "WD" plus "accretion disk" (which is now commonly done in reports of ultraviolet spectra). Our *opinion* is that WD light in these CVs is predominantly just the heat left over from time-averaged accretion – in which case the separation is unwarranted, and theoretical bolometric corrections from WD or disk atmospheres (or better yet, empirical corrections from UV-optical-IR flux distributions) are appropriate. But real evidence on this point is still lacking. Readers holding a different opinion, or disturbed by these uncertainties, may prefer the estimate from $M_v$ (if they are still reading).

If so, it should have a discernible effect on the entire population of short-period CVs, not just on recent novae. We can study that population in several ways.

### 11.1.1  Minimum $P_{orb}$, $q(P_{orb})$, $R_2(P_{orb})$, and $R_2(M_2)$

A useful diagnostic diagram for short-period CVs is $q(P_{orb})$; $P_{orb}$ is easily and accurately learned from spectroscopic and photometric observations, and $q$ is often learned from the fractional period excess of superhumps (and occasionally from direct dynamical observation). This yields distributions like Figure 8, which is an updated version of Figure 6 of P11. The solid curve is the theoretical expectation, based on the assumption of evolution driven purely by GR, a 0.75 Mo WD, and secondaries which start as 0.2 Mo main-sequence stars and then evolve as their thermal timescales increase and ultimately exceed their mass-loss timescales (causing a minimum $P_{orb}$, followed by "period bounce" as binary evolution continues). Figure 8 shows a pronounced disagreement with theory. One aspect of this disagreement has been much discussed: minimum period actually occurs at ~80 min, compared to the theoretical 70 min (Patterson 2001, King et al. 2002, Barker & Kolb 2003, Gaensicke et al. 2009). But this is actually a special case of a more general disagreement: at every value of $q$, the measured values of $P_{orb}$ appear to be too high.

The same is true of $R_2(P_{orb})$ and $R_2(M_2)$. Prior to period bounce, the donor stars are somewhat larger than they would be on the "main sequence" (Figure 2 of Patterson 2001; Figures 10 and 11 of P05; Figures 4, 9, and 10 of KBP).[42]

---

41 This is roughly probed by the values of $q$ (=$M_2/M_1$) suggested by the superhump period excesses, which are consistent with the normal $q(P_{orb})$ relation (P05).
42 These various relations come mainly from the superhump period excesses, and are substantially equivalent. Roche-lobe geometry implies that $R_2 \sim (P_{orb})^{2/3} M_1^{1/3} q^{1/3}$

From the earliest presentations of such figures, it has been recognized that the disagreement can be remedied by a boost in the angular-momentum loss by a factor of ~3 over that of pure GR (Patterson 1998, 2001; King et al. 2002; P05; and especially KBP, who calculate that the boost should be a factor of 2.47+-0.22). This boost is artificial – not required or suggested by any known physics. The role of enhanced $\dot{J}$ is primarily to shrink the binary dimensions at a faster rate; this increases the donor star's mass loss and forces the star out of thermal equilibrium earlier in its evolution. Thus the donor is whittled away faster, and minimum $P_{orb}$ is reached sooner. This "2.47 GR" prescription produces evolution along the dashed curve in Figure 8 (calculated from Table 3 of KBP).

### 11.1.2 The White Dwarf's $T_{eff}(P_{orb})$

Another probe of evolution is provided by the distribution of WD $T_{eff}$ versus $P_{orb}$. Every gram of accreted matter heats the WD by accretion and compression, so perhaps we can measure accretion rates by measuring $T_{eff}$. The quantitative basis for this was presented by Townsley & Bildsten (2004), and was warmly welcomed, on the grounds that $T_{eff}$ automatically averages over very long time intervals – roughly the thermal time scale of the WD's envelope. Summarized in Figure 5 of Townsley & Gaensicke (2009), the results show an average $T_{eff} \approx 15000$ K for the short-$P_{orb}$ stars, compared to ~12000 K expected if GR is the sole driver of mass transfer. At face value, that could be another clue signifying an additional driver (such as postnova heating). However, the method has several problems. First, an accurate $T_{eff}$ measurement requires an observation (usually a spectrum) in the vacuum ultraviolet, and therefore is difficult to obtain. Only ~50 such measurements have been made in the 30 years since the relevant telescopes have been available (HST, IUE, Galex), and only ~15 sample the regime of interest (nonmagnetic, short-$P_{orb}$). So the observational basis is still sparse. Second, we know from repeated measurements of the best-studied system that the WD cools substantially in the long aftermath of dwarf-nova outbursts (WZ Sge: Figure 6 of Godon et al. 2006). Since outbursts of lesser-known systems are easily missed, this casts some doubt on the "long-term average" advantage. Third, the predicted $T_{eff}$ strongly depends on WD mass ($M^{1.7}$), as well as $\dot{M}$; that significantly weakens the constraint on $\dot{M}$. And fourth, the "observed" $T_{eff}$ is usually a best value from a (log $g$ – $T_{eff}$) grid, which implies another dependence on WD mass. At present, these issues appear to make the method less useful than

---

[Eq. (10) of P05]; so when $P_{orb}$ is "too large" for a given $q$, then $R_2$ is also too large ("bloated"). We prefer the $q(P_{orb})$ version, since it is basically a relation between measured quantities; but $R_2(P_{orb})$ and $R_2(M_2)$ convey more physical insight, since they quantitatively and *visually* show the secondary's departure from thermal equilibrium.

$q(P_{orb})$ and its cousins.

### 11.1.3  $M_v(P_{orb})$, and the Second Parameter in CV Evolution

We understand now that $P_{orb}$ is the main determinant of a CV's luminosity; this is proved by Figure 7 of P84, and is assumed by all theories of CV evolution. But at least for short-$P_{orb}$ stars, there is a great deal of scatter in luminosity at fixed $P_{orb}$. This is shown by Figure 9, which is an updated version[43] of P11's Figure 6. This scatter is very surprising! Except near minimum period, Figure 8 showed that $q$, and therefore $M_2$, is fairly well determined by $P_{orb}$; and CV secondaries obey a well-defined mass-radius relation (Figure 12 of P05, Figure 3 of K06; Figure 4 of KBP). Why should stars of the same mass and radius transfer matter at greatly different rates? For example: BK Lyn, Z Cha, and RZ Leo are all dwarf novae, have similar $q$ values, and have orbital periods within 1% of each other. All are well separated from the period-bounce regime.[44] Yet their luminosities differ by a factor of ~70 (estimated by P11 as $<M_v>$ = 5.7, 9.1, and 10.3 respectively). Why should stars of identical $P_{orb}$ differ by a factor of 70 in accretion rate?

One possibility is that some low-mass secondaries manage to retain some fraction of the magnetic braking which drove their mass transfer when $M_2$ was higher. That's a solution we advocated previously (Patterson 2001, and implicitly also in KBP), and it might be true[45]. But it has some demerits. It's a tad *deus ex machina;* it requires that extra magnetic braking be idiosyncratically allotted to one star rather than another of the same mass; and it creates a puzzle of how, for example, BK Lyn could have ever reached the same orbital period as RZ Leo. With angular-momentum loss 70x greater, BK Lyn should have "bounced" at a much longer period.

Another possibility is mass-transfer cycles. Many papers (e.g. King et al. 1996, 1997) have studied how irradiation of the donor star can produce cyclic radius variations, and therefore cycles in the rate of mass transfer. As long as the cycles are long compared to our observation span (~100 years), this can produce a large spread in $\dot{M}$ at a given P$_{orb}$, as desired.  These may well exist and be observable, but their relation to the BK Lyn/ER UMa puzzles is not clear – because the latter stars represent only 1-2% of CVs.  Such rarity constrains the cycles to be very asymmetric (more off than on, by a factor 50-100).

---

43 Now including nova remnants and the ER UMa stars, which were excluded by P11 as "anomalous" (and speculated to be the result of nova heating).
44 Where low luminosities are likely caused by a different effect: simply low $M_2$ [Mdot scales roughly as $(M_2)^2$ for low $M_2$, if the driver is pure GR]. A good example pair is FO And and GD 552; they differ in $P_{orb}$ by just 0.4 percent, but in $<M_v>$ by >4 mag (P11 Table 2). GD 552 is the poster-child period-bouncer.
45 To some degree. In order to produce the known 2.2-2.8 hr period gap, the shutoff of strong magnetic braking must be common, large, and sudden. But not necessarily *equally* large and sudden for every star; the few oddballs (in-the-gap, or bright stars at short period) suggest, or at least permit, some variance in this process.

The million-year nova cycle appears to have that property.  That version of the "cycles" hypothesis – the one we like now – assumes that stars are seen at various stages on their decline light curves, rather than being fully subsided from nova eruptions ("quiescence").  Figure 9 separately identifies the novae, BK Lyn, and the known ER UMa stars, in order to illustrate the idea.  Nova remnants shine with M$_v$=+4 about 30 years after outburst, then fall vertically downward, with ever-increasing slowness, according to the decline suggested by Figure 7.  This explains the ER UMa stars, and also sprinkles the general population with many stars still showing, to a lesser degree, the hangover from their most recent nova event.  Hence there is a second parameter important for CVs: **time since the last nova eruption**.  The scatter in Figure 9 can then be explained, without need to explicitly invoke magnetic braking.[46] Instead, the candidate culprit is irradiation of the secondary.  Since the nova's million-year cycle time is short compared to the secondary's thermal timescale, the effect of irradiation would be mainly cumulative (expanding the star by ~20%[47]) but also somewhat short-term – expanding the radius by a few pressure scale heights, responding to changes in illumination (on the timescale of postnova cooling).

The physical idea here is that irradiation blocks the secondary's own outward flux, since a star relies on its *dT/dr* gradient in order to radiate.  The star then expands to re-establish that gradient, and $\dot{M}$ consequently increases.  This occurs in cycles, and hence a given star can be found in a relatively high- $\dot{M}$ state ("novalike variable") or relatively low ("dwarf nova").  As the source of the irradiation, the several previous studies have invoked combinations of WD, boundary-layer, and accretion-disk – with appropriate corrections for their geometry (e.g. the accretion-disk being probably unimportant, since it is flat and does not radiate towards the secondary).

These effects may well be greater in a short-period secondary.  By the standards of short-$P_{orb}$ CVs, the eruption and long postnova effects envisioned in Figure 7 generate a lot of light. Integration under the full light curve, from explosion to quiescence, suggests that $3 \times 10^{45}$ ergs are radiated during the first 1-2 years, followed by another $3 \times 10^{45}$ ergs during the next 20000 years, and another $2 \times 10^{45}$ ergs in the next 500000 years. The secondary's own radiation amounts to just $2 \times 10^{44}$ ergs during this interval, and thus is outshone by a factor ~40. We estimate that ~3% of the radiated light falls on the secondary, and therefore the perturbation on the secondary's structure could be significant.[48]

---

46 However, to effectively drive evolution, there must be some angular-momentum loss; otherwise, mass transfer from a low-mass secondary, conserving angular momentum, will widen the binary and thereby quench $\dot{M}$ .  This has received the label "consequential" angular-momentum loss (CAML) – proportional to the extra $\dot{M}$ – and is necessary to maintain irradiation-driven cycles (King et al. 1996).  We don't know what that mechanism is, but one possibility is a stellar wind from the donor.  The donor's physical circumstances seem promising for a wind (high specific angular momentum, low effective gravity, fast rotation, convection, inverted *dT/dr*).  Another is frictional angular-momentum loss (FAML, MacDonald 1986) from the donor, orbiting in the nova's wind.

47 This is the right amount to account for the "bloating" which underlies the disagreement with the GR prediction in Figure 8.

### 11.1.4  Plus, While We're at it, Maybe a Third

Figure 9 emphasizes that recent novae and ER UMa stars are quite far from their natural home. They're much too bright, and we blame that on long-term heating effects from a recent nova eruption. Figure 10 is an attempt to learn, and understand, just what is that natural home. Figure 10 compares the empirical $M_v(P_{orb})$ curve, excluding recent-nova suspects, with two theories[49] of evolution: driven by GR, and driven by angular-momentum loss 2.47x greater (the best fit in KBP's analysis). If our excuse for the outlandish behavior of the ER UMas is accepted, then Figures 8 and 10 comprise the basic data which an evolution theory needs to explain.

Both figures show that GR is a poor fit: it predicts stars which are too faint, with $q$ too large and a minimum $P_{orb}$ too small. The KBP fit is far superior for $P_{orb}>0.06$ d. But it appears to predict a minimum $P_{orb}$ slightly too long, and perhaps overestimates[50] the brightness of stars near and past period bounce.  GR certainly seems to need a boost, but the physical nature and mathematical description of that boost still elude us.

---

48 The geometrical dilution factor alone is ~3%, assuming point-source illumination from the WD. Disk flatness, disk shadowing, and the secondary's albedo will lower it. But this fraction is likely to be higher in short-period binaries, for several reasons. During the active nova phase, there is likely no disk to shadow the very hot WD. During the BK Lyn/ER UMa phase, there is a potentially shadowing disk, but the shadow is probably mitigated or negated by disk tilt (evidenced by the negative superhump – which, based on its characteristic detection among the ER UMas, is probably an enduring feature). And during the long "true quiescence", the disk is optically thin and is effectively neither flat nor shadowed; this is proved by the presence of sharp, deep eclipses of the WD in suitably inclined systems (which attest to the clear lines of sight between WD and secondary). We think that ~3% average irradiation is a pretty good guess.

49 We calculate $M_v$ from the theoretical $\dot{M}$ by using $L_{bol} = GM\dot{M}/2R$ and then applying a bolometric correction of 1.4 mag, which we measure from observations of the best-studied dwarf novae in outburst (U Gem, SS Cyg, WZ Sge). Broadband colors of dwarf novae in outburst are remarkably close to a standard (*B-V*=0, *U-B*=-0.7), suggesting that the disk's bolometric corrections are not far from a standard value.

## 12. NOVAE THROUGH THE CENTURIES

What about all the short-$P_{orb}$ novae that erupted between the years 101 and 1942?  Six such stars have erupted since 1942, suggesting a rate of ~0.1/yr.  That suggests the existence of ~180 short-period novae, all quite youthful and shining as novalike variables with $M_V$~+5-6 according to Figure 7.  Are any of these stars known?

According to the main argument motivating this research ("BK Lyn is the only short-period novalike in the sky"), no... and that would seem to be embarrassing for the view peddled here.  But *distance* may play a big role.  The rate-setting collection of six stars is at a mean distance of ~3 Kpc; this implies a distance modulus of 12.3, and probably ~1 magnitude of absorption for average lines of sight.  These putative stars should therefore be mostly found with $V$~17-19, depending on youth, distance, and galactic latitude.  Few surveys would flag such stars.  They will hide from some surveys by virtue of their minor variability ("novalike") and weak emission lines – and from others by virtue of faintness and/or interstellar reddening.  The only large survey which would have a good chance of detecting these stars is the Sloan Digital Sky Survey, which mainly finds, and obtains useful spectra of, stars with $V$~15.5-20.  Up to the present, the SDSS has identified ~400 CVs, and our inspection of that roster (contained in the series of papers starting with Szkody et al. 2002) shows ~10-20 which could be considered as candidates.  It would be fascinating to study these stars more thoroughly, with postnova credentials in mind.  Although the SDSS is not ideal since it favors high galactic latitudes, we predict that a few will be found – the youthful remnants of novae over the last 2000 years.

It would also be worthwhile to snoop among the known SW Sex stars.  These have mysteriously high temperatures, brightnesses, and likely accretion rates; so they're definitely prime suspects.

## 13. SUMMARY

1. Seasonal light curves of BK Lyn demonstrate that somewhere in the interval 2002-5, this novalike variable mutated into a card-carrying dwarf nova of the ER UMa class, with superoutbursts occurring every 45±3 days.  The light curve *shape* attests to this, and so does the pattern of positive superhumps, which is essentially identical to that of dwarf novae: sudden creation at large amplitude at the onset of superoutburst, and then slow decay as the star declines. The period is 4.6% greater than $P_{orb}$, and somewhat unstable – also typical of dwarf novae.

---

50 Far from certain.  Stars at short $P_{orb}$ have coarser observational constraints, mainly because they are intrinsically fainter and seldom erupt.

2. The star has shown negative superhumps, with a period 3.0% shorter than $P_{orb}$, in every season since 1999[51]. These are much more stable. Such a phenomenon is fairly common in novalike variables; but among dwarf novae, it exists only in (some) ER UMa stars. The amplitude and phase of this signal appear to pay no attention to the dwarf-nova outbursts. Accompanying the negative superhump is a signal with a period of 2.54±0.03 days – exactly at the beat frequency between orbit and negative superhump, and therefore likely to be the disk's actual wobble period.

3. It is not our intention here to explore in detail the interpretation of superhumps, but we adopt the now-common view that positive and negative superhumps arise respectively from apsidal advance and nodal regression in an eccentric and tilted accretion disk (Harvey et al. 1995, Patterson et al. 1997, Patterson 1999, Skillman et al. 1999, Wood & Burke 2007, Montgomery 2012). If the apses advance with a frequency A, and the nodes regress with a frequency N, then BK Lyn shows signals at these frequencies: N, ω-A, ω, ω+N, etc. In fact, nearly all superhumping CVs (>100 of them) obey these rules: all apparitions of A are in the form -A, all apparitions of N are in the form +N, all negative superhumps are accompanied by an apparition of N; and no positive superhumps are accompanied by an apparition of A.

4. The dwarf-nova standard-candle relation implies a distance of about 800 pc. This implies that the time-averaged $M_v$ is now +5.7, which translates to $M_{bol}$~+3.7, or L~$1.5\times10^{34}$ erg/s. Supplied by accretion onto a 0.8 Mo white dwarf, this implies a present-day accretion rate ~$2\times10^{-9}$ Mo/yr. This is roughly the theoretical limit between novalikes and dwarf novae. There's a decent chance that BK Lyn will temporarily mutate back into a novalike in the near future – which would then earn it a new award: "the first short-period Z Cam star".

5. Arguments from positional coincidence, galactic latitude, absolute magnitude, WD temperature, nova physics, and the rarity (uniqueness... and, as of 2005, nonexistence) of short-period novalikes suggest that BK Lyn is the old-nova remnant of the "guest star" of 101 A.D.

6. This suggests the following timescales for a short-period CV's relaxation following a classical-nova eruption: 2000 years as a novalike, 15000 years as an ER UMa dwarf nova (gradually declining to an ordinary SU UMa), and 500,000 years in true

---

[51] Although we do not rule out their presence in 1992-4. Those earlier campaigns did not span a wide range of terrestrial longitude. They were sufficient to establish that the main signal was the apsidal superhump – but not to exclude the presence of a (weak) concomitant negative superhump, separated by ~1.00 c/d.

quiescence.

7. Whether or not BK Lyn is actually Nova Lyn 101, this hypothesis – of very lengthy postnova relaxation – can take us a long way. It can explain:
   (a) why ER UMa stars exist (because they are remnants of recent novae);
   (b) why they're rare (because that phase is only 2% of a nova's full eruption cycle);
   (c) why short-period novalikes are even rarer (because that phase is even briefer);
   (d) why historical short-period novae always decline to brightness states far above true quiescence (because their cooling clocks are just getting started, a la Figure 7);
   (e) why CVs show a large spread of $<M_v>$ and white-dwarf temperature at a fixed $P_{orb}$, contrary to the expectation based on a pure-GR driver of evolution (in part, because the stars have not fully cooled to quiescence);
   (f) why the minimum $P_{orb}$ among hydrogen-rich CVs is too long to be consistent with a simple GR model (because the extra mass transfer induced in the long postnova phase drives the secondary out of thermal equilibrium faster); and perhaps
   (g) why post-period-bounce stars are hard to find (because evolution proceeds somewhat faster than predicted by GR, thus burning them out quickly).

8. To explain this lengthy postnova relaxation, we invoke irradiation of the secondary. We appeal to this both for its long-term effect [e.g. to explain the puzzles contained in the $q(P_{orb})$ distribution] and for briefer phenomena interpreted as large postnova effects (explaining oddities like BK Lyn and the ER UMas in Figure 9). Actually, it might explain the *full* range of $<M_v>$ variation at a fixed $P_{orb}$. But the last is admittedly a stretch; it may be that after removing the ER UMa class from the upper parts of Figure 9, and the period bouncers from the lower parts, there is no effect left to explain. (In other words, the remaining scatter might simply arise from observational error, dispersion in WD mass, etc.). This irradiation can come initially from the hot WD, and later from the disk itself (significantly aided by its non-coplanar orientation). Detailed calculation of the secondary's response is needed to explore these possibilities.

9. For a high WD mass, the effects hypothesized here become more extreme. The short eruption cycle predicted for high $M_1$ (by the tables of Yaron et al. 2005, and all models of classical novae) implies that the secondary can never reach quiescence; a few decades or centuries of elevated postnova accretion could suffice to trigger a new eruption. This is the T Pyx scenario ("assisted stellar suicide", Knigge et al. 2000). In addition to the strong dependence on $M_1$, it's also likely to favor low $M_2$, since the heating can then greatly overwhelm the secondary's own luminosity, leading to its actual expansion (not merely inability to contract fast enough to preserve thermal equilibrium in the presence of mass loss). This may be why T Pyx has no close relatives – it requires not only high $M_1$, but also short $P_{orb}$, which is rare among novae.

10. Several weaknesses in the argument remain (plus any we haven't recognized).

    (a) Several lines of evidence suggest that BK Lyn's oddity arises from a short-lived postnova phase of evolution, but none is compelling. It would be mighty nice to find a proof (a nova shell, perhaps?).

    (b) The calculation of postnova light compared to quiescent GR-driven light is rough; if the timescale for postnova cooling is faster than we have estimated, then there may be little time-averaged effect on the star's brightness. This might leave undisturbed the explanation of ER UMa stars and the scatter in $M_v(P_{orb})$, but would then require a separate hypothesis ("residual magnetic braking" or some other angular-momentum loss mechanism) to explain the oddities of Figures 8 and 10.

    (c) Theoretical study of the secondary's response to a radiation bath, of the type we hypothesize, is needed. Is it really true that the main effect would be long-term heating, rather than prompt re-emission from the heated hemisphere?

    (d) We may not have correctly distinguished cause and effect. The ER UMas are distinctive for their high luminosity and hot WDs, and we blame that on irradiation by a recent nova. But they're also distinctive for their apparently tilted disks, which could effectively irradiate the secondary and thereby cause the enhanced mass transfer. Which is the underlying cause? We advocate the recent nova, on two grounds: the probable identity of BK Lyn with an ancient nova, and the simple seduction of numbers ($10^{38}$ erg/s trumping $10^{34}$ erg/s). But that might be wrong... and it might be a combination.

    (e) It would be nice to find some other ancient-nova candidates among short-period CVs – or, better yet, find nova shells around these or the existing ER UMas.

11. This research was supported by grants from the National Science Foundation (AST-0908363 and AST-1211129) and the Mount Cuba Astronomical Foundation. We benefited from discussions with Brad Schaefer, Christian Knigge, Patrick Wils, Jeno Sokoloski, and Jim Applegate. Nearly all the data was acquired in suburban backyards, with midget telescopes (~0.3 m) operated by people with "day jobs". We're less sure about the amateur status of whoever who did the hard work 2000 years ago, but... Vive le citizen science!

REFERENCES

Antonyuk, O.I. & Pavlenko, E.P. 2005, The Astrophysics of Cataclysmic Variables and Related Objects ed. J.M. Hameury & J.P. Lasota, ASP Conf. Ser. 330, 379.
Baraffe, I. & Kolb, U. 2000, MNRAS, 318, 354.
Barker, & Kolb, U. 2003, MNRAS, 340, 623.
Barrett, P., O'Donoghue, D., & Warner, B. 1988, MNRAS, 233, 759.
Bianchini, A. et al. 2012, A&A, 539, A94.
Cannizzo, J.K. & Kenyon, S.J. 1986, ApJ, 309, L43.
Cannizzo, J.K et al. 2012, ApJ, 747, 117.
Chochol, D. et al. 1997, A&A, 318, 908.
Clark, D.H. & Stephenson, F.R. 1977, The Historical Supernovae (Pergamon: New York).
Collazzi, A.C., Schaefer, B.E., Xiao, L., Pagnotta, A., Kroll, P., Lochel, K., & Henden, A.A. 2009, AJ, 138, 1846.
Coyne, R. et al. 2012, MNRAS, in press.
de Miguel, E. et al. 2012, SASS, 31, 79.
Diaz, M.P. & Steiner, J.E. 1994, ApJ, 425, 252.
Diaz, M.P. et al. 1995, MNRAS, 277, 959.
Duerbeck, H.W.1992, MNRAS, 258, 629.
Gaensicke, B.T. et al. 2009, MNRAS, 397, 2170.
Gao, W. et al. 1999, ApJ, 527, L55.
Godon, P., Sion, E.M., Cheng, F., Long, K.S., Gaensicke, B.T., & Szkody, P. 2006, ApJ, 642, 1018.
Green, R.F., Schmidt, M., & Liebert, J. 1986, ApJS, 61, 305.
Green, R.F., Ferguson, D.H., Liebert, J.E., & Schmidt, M.1982, PASP, 94, 560.
Hachisu, I. & Kato, M. 2007, ApJ, 662, 552.
Hachisu, I. & Kato, M. 2010, ApJ, 709, 680.
Hachisu, I., Kato, M., & Cassatella, A. 2008, 687, 1236.
Harvey, D.A., Skillman, D.R., Patterson, J., & Ringwald, F. 1995, PASP, 107, 551.
Hellier, C. 2001, PASP, 113, 469.
Hertzog, K.P. 1986, Observatory, 106, 38.
Honeycutt, R.K., Robertson, J.W., & Kafka, S. 2011, AJ, 141, 121.
Ho, P.Y. 1962, Vistas in Astronomy, 5, 127.
Hsi, T.-T. 1958, Smithson. Contr. Astrophys., 2, 109.
Ishioka, R. et al. 2001, PASJ, 53, 51.
Kato, T. et al. 2002, PASJ, 54, 1029.
Kato, T. et al. 2003, MNRAS, 341, 901.
Kato, T. et al. 1999, in Disk Instabilities in Close Binary Systems, ed. S. Mineshige & J.C. Wheeler (Tokyo: Universal Academy Press), p. 45.
Kato, T. et al. 2009, PASJ, 61, S395-616.
Kato, T. et al. 2010, PASJ, 62, 1525.
King, A.R., Frank, J., Kolb, U., & Ritter, H. 1995, ApJ, 444, L37.
King, A.R., Frank, J., Kolb, U., & Ritter, H. 1996, ApJ, 467, 761.
King, A.R., Schenker, K. & Hameury, J.M. 2002, MNRAS, 335, 513.

Knigge, C. 2006, MNRAS, 373, 484 (K06).
Knigge, C., King, A.R., & Patterson, J. 2000, A&A, 364, L75.
Knigge, C., Baraffe, I., & Patterson, J. 2011, ApJS, 194, 28 (KBP).
Kolb, U. & Baraffe, I. 1999, MNRAS, 309, 1034.
Livio, M. & Shara, M.M. 1987, ApJ, 319, 819.
Livio, M., Govarie, A., & Ritter, H., A&A, 246, 84.
Lubow, S.H. 1991, ApJ, 381, 268.
Lubow, S.H. 1992, ApJ, 398, 525.
MacDonald, J. 1986, ApJ, 305, 251.
McLaughlin, D.B. 1960, in Stellar Atmospheres, ed. J.L. Greenstein (Chicago: U. Chicago), p. 585.
Montgomery, M. M. 2009, ApJ, 705, 603.
Montgomery, M. M. 2012, ApJ, 753, L27.
Nelson, T. et al. 2012, arXiv 1211.3112.
Nogami, D. et al. 2003, A&A, 404, 1067.
O-Donoghue, D. et al. 1989, MNRAS, 240, 41.
Ohshima, T. et al. 2012, PASJ, 64, L3.
Olech, A. et al. 2004, AcA, 54, 57.
Olech, A. et al. 2007, AcA, 57, 331.
Olech, A. et al. 2008, AcA, 58, 131.
Olech, A., Rutkowski, A., & Schwarzenberg-Czerny, A. 2009, MNRAS, 399, 465.
Orio, M., Nelson, T., Bianchini, A., Di Mille, F. & Harbeck, D. 2010, ApJ, 717, 739.
Osaki, Y. 1995, PASJ, 47, L11.
Osaki, Y. 1996, PASP, 108, 39.
Otulakowska-Hypka, M., Olech, A., de Miguel, E., Rutkowski, A., Koff, R., & Bawkowska, M. 2012, MNRAS, in press.
Patterson, J. 1984, ApJ, 54, 443 (P84).
Patterson, J. 1999, in Disk Instabilities in Close Binary Systems, ed. S. Mineshige & J.C. Wheeler (Kyoto: Universal Academy Press), p. 61.
Patterson, J. 1998, PASP, 110, 1132.
Patterson, J. 2001, PASP, 113, 736.
Patterson, J. 2011, MNRAS, 411, 2695 (P11).
Patterson, J. & Warner, B. 1998, PASP, 110, 1026.
Patterson, J. et al. 1995, PASP, 107, 1183.
Patterson, J. et al. 1997, PASP, 109, 468.
Patterson, J. et al. 1998, PASP, 110, 380.
Patterson, J. et al 2002, PASP, 114, 721.
Patterson, J. et al. 2003, PASP, 115, 1308.
Patterson, J. et al. 2005, PASP, 117, 1204 (P05).
Patterson, J. et al. 2012, in preparation.
Pavlenko, E. et al. 2010, 17[th] European White Dwarf Workshop, AIP Conf. Proc. 1273, 320.
Pskovskii, Y.P. 1972, Sov. Astr., 16, 23.
Rappaport, S.A., Joss, P.C., & Webbink, R.L. 1982, ApJ, 254, 616.
Retter, A. & Lipkin, Y. 2001, A&A, 365, 508.

Richman, H.R., Applegate, J.H., & Patterson, J. 1994, PASP, 106, 1075.
Ringwald, F.A., Thorstensen, J.R., Honeycutt, R.K., & Robertson, J.W. 1996, MNRAS, 278, 125.
Robertson, J.W., Honeycutt, R.K., & Turner, G.W. 1995, PASP, 107, 443.
Robinson, E.L. 1975, AJ, 80, 515.
Rutkowski, A. et al. 2009, A&A, 497, 437.
Schaefer, B.E. 2005, ApJ, 621, L53.
Schaefer, B.E. & Collazzi, A. 2010, AJ, 139, 1831 (SC).
Schaefer, B.E. et al. 2012, arXiv: 1109.0065.
Schaefer, B.E. et al. 2012, private communication.
Schreiber, M.R. & Gaensicke, B.T. 2001, A&A, 375, 937.
Schreiber, M., Gansicke, B.T. & Cannizzo, J.K. 2000, A&A, 362, 268.
Shanley, L., Ogelman, Gallagher, J.S., Orio, M., & Krautter, J. 1995, ApJ, 438, L95.
Shara, M.M. et al. 1984, ApJ, 278, 845.
Shara, M.M. et al. 2007, Nature, 446, 159.
Shara, M.M. et al. 2012, ApJ, 756, 107.
Sirotkin, F.V. & Kim, W. 2010, ApJ, 721, 1356.
Skillman, D.R. & Patterson, J. 1993, ApJ, 417, 298 (SP).
Skillman, D.R., Harvey, D., Patterson, J., & Vanmunster, T. 1997, PASP, 109, 114.
Skillman, D.R. et al. 1999, PASP, 111,1281.
Smak, J.I. 1989, AcA, 39, 317.
Somers, M.W., Mukai, K. & Naylor, T. 1996, MNRAS, 278, 845.
Still, M. et al. 2011, ApJ, 717, L113.
Szkody, P. et al. 2002, AJ, 123, 430.
Tamburini, F. et al. 2007, A&A, 464, 697.
Thorstensen, J.R. et al. 1997, PASP, 109, 477.
Thorstensen, J.R. & Taylor, C.J. 2000, MNRAS, 312, 629.
Townsley, D.M. & Bildsten, L. 2004, ApJ, 600, 390.
Townsley, D.M. & Bildsten, L. 2005, ApJ, 628, 395.
Townsley, D.M. & Gaensicke, B.T. 2009, ApJ, 693, 1007.
Urban, J.A. & Sion, E.M. 2006, ApJ, 642, 1029.
Uthas, H., Knigge, C. & Steeghs, D. 2010, MNRAS, 409, 237.
Warner, B. 1988, Nature, 336, 129.
Whitehurst, R. 1988, MNRAS, 232, 35.
Wood, M.A. & Burke, C.J. 2007, ApJ, 661, 1042.
Wood, M.A. et al. 2011, ApJ, 741, 105.
Yaron, O., Prialnik, D., Shara, M.M., & Kovetz, A. 2005, ApJ, 623, 398.
Zellem, R. et al. 2009, PASP, 121, 942.

Table 1

Summary Observing Log

| Observer | CBA Station | Year nights/hours | | | | |
|---|---|---|---|---|---|---|
| | | 1999 34/225 | 2002 42/306 | 2005 19/94 | 2011 58/324 | 2012 215/1108 |
| Tom Krajci | New Mexico | – | – | – | | 33/214 |
| David Cejudo | Spain (Madrid) | – | – | – | | 24/153 |
| Enrique de Miguel | Spain (Huelva) | – | – | – | 13/63 | 37/164 |
| Tut Campbell | Arkansas | – | – | – | 11/44 | 27/148 |
| Shawn Dvorak | Orlando | – | – | – | 4/15 | 22/101 |
| Josch Hambsch | Belgium (Mol) | – | – | – | 21/126 | 12/64 |
| John Rock | England (Wilts) | – | – | – | – | 11/52 |
| David Boyd | England (Oxford) | – | – | – | – | 7/26 |
| Etienne Morelle | France | – | – | – | – | 10/62 |
| Joe Ulowetz | Illinois | – | – | – | – | 15/63 |
| Richard Sabo | Montana | – | – | – | – | 10/31 |
| Others | Various | – | – | 6/30 | – | 3/15 |
| Arto Oksanen | Finland | – | – | 5/15 | – | 4/17 |
| Gianluca Masi | Italy | – | – | – | 1/3 | – |
| Mike Potter | Baltimore | – | – | – | 3/14 | – |
| Tonny Vanmunster | Belgium (Landen) | 2/7 | – | 7/36 | – | – |
| Anthony Kroes | Wisconsin | – | – | 4/19 | – | – |
| Brian Martin | Alberta | – | 9/61 | 2/9 | – | – |
| Jerry Foote | Utah | – | 17/145 | – | – | – |
| David Skillman | East (Maryland) | 7/40 | 10/67 | – | – | – |
| Robert Fried | Flagstaff | 8/47 | 4/25 | – | – | – |
| Jonathan Kemp | MDM | 2/8 | 2/8 | – | – | – |
| David Harvey | West (Tucson) | 8/80 | – | – | – | – |
| Lasse Jensen | Denmark | 7/43 | – | – | – | – |

Table 2

Frequencies in Years of Observation (cycles/day)

| Signal Identification | Year | | | | | |
|---|---|---|---|---|---|---|
| | 1992 (±0.002) | 1993 (±0.001) | 1999 (±0.009) | 2002 (±0.002) | 2005 (±0.006) | 2012* (±0.003) |
| N | – | – | 0.400 | 0.393 | | 0.395 |
| ω-A | 12.7335 | 12.7280 | 12.770 | 12.760 | 12.752 | |
| ω | | | | 13.344 | | |
| ω+N | | | 13.738 | 13.738 | 13.740 | 13.728 |
| 2(ω-A) | | | 25.561 | 25.521 | 25.506 | 25.543 |
| 2(ω+N) | | | 27.460 | 27.470 | 27.471 | |
| 3(ω+N) | | | | 41.215 | | |
| 4(ω-A) | | | 51.07 | 51.041 | | |
| 4(ω-A)+N | | | | 64.759 | | |

\* Many weaker signals (probable sidebands and harmonics) are seen during 2011-12, but secure measurement and identification is dependent on how the large dwarf-nova brightness changes are removed.

Table 3

Maximum Light* (HJD 2,455,000+) of Negative Superhump

| | | | | | |
|---|---|---|---|---|---|
| 928.7875 | 929.8840 | 930.9010 | 931.9776 | 952.5410 | 952.7598 |
| 953.8545 | 955.6062 | 955.8964 | 956.8442 | 957.7880 | 958.4500 |
| 958.8903 | 959.7602 | 960.7104 | 963.8430 | 964.5710 | 966.5374 |
| 967.4817 | 968.6458 | 968.7902 | 969.6650 | 970.5378 | 970.7574 |
| 971.7734 | 983.5046 | 983.7894 | 984.5242 | 984.7344 | 985.4683 |
| 986.5610 | 986.7774 | 987.4417 | 987.7242 | 988.6713 | 989.5438 |
| 989.7590 | 995.4400 | 997.4133 | 998.4939 | 999.5180 | 999.7334 |
| 1000.4658 | 1000.6754 | 1001.4832 | 1001.6298 | 1002.5074 | 1003.7432 |
| 1005.5016 | 1006.4467 | 1007.4705 | 1008.4190 | 1009.4373 | 1009.7353 |
| 1010.5348 | 1010.6810 | 1011.4806 | 1011.6956 | 1012.4273 | 1013.5202 |
| 1013.6664 | 1014.5428 | 1015.4900 | 1015.6356 | 1016.4352 | 1017.4550 |
| 1018.3995 | 1053.732 | 1054.750 | 1055.7735 | 1056.721 | 1058.757 |
| 1059.774 | 1060.721 | 1076.745 | | | |

\* Each timing is averaged over 2-5 superhump cycles.

Table 4

Roster of ER UMa Stars ($T_{super}$<120 d) and Short-Period Novae***

| Star | $P_{orb}$ (d) | $T_{super}$ (d) | $T_{normal}$ (d) | V | Distance[†] (pc) | $<M_v>$[†] | $T_{wd}$ | References |
|---|---|---|---|---|---|---|---|---|
| RZ LMi | ~0.058* | 19 | 4 | 14-16.7 | 700 | 7.0 | 33000 | 1, 9, 10, 20 |
| BK Lyn[§] | 0.0750 | 45 | 5 | 14-16 | 800 | 5.7 | 35000 | 1, 2, 3 |
| ER UMa[§] | 0.0637 | 44 | 5 | 12.7-15 | 350 | 6.7 | 21000 | 1, 5, 6, 7, 20 |
| V1159 Ori | 0.0622 | 48 | 4 | 12.8-15.3 | 370 | 7.2 | 20000 | 1, 2, 7, 8, 20 |
| DI UMa | ~0.0546* | 32 | 7 | 14.7-17.6 | 800 | 7.0 | >20000 | 1, 11, 12 |
| IX Dra | ~0.0665* | 58 | 3-4 | 15-17.1 | 800 | 7.1 | | 1, 16, 17, 31 |
| MN Dra[§] | ~0.099 | 70 | 12 | 15.7-19.8 | 1100 | 7.3 | | 1, 15, 4 |
| BF Ara[§] | 0.0842 | 83 | – | 14-18 | 500 | 6.9 | | 1, 18, 19 |
| V503 Cyg[§] | 0.0777 | 90 | 6 | 13-17.5 | 430 | 7.5 | | 1, 13, 14 |
| V1504 Cyg | 0.0695 | 112 | 11 | 13.8-17.5 | 850 | 7.3 | | 34, 35, 36, 37 |
| V344 Lyr[§] | ~0.087* | 118 | 18 | 15-19 | 800 | 7.5 | | 38, 32, 31 |
| CP Pup | ~0.0614* | Nova 1942 | | 15.5# | 900 | 5.2 | | 21, 22, 30 |
| RW UMi | ~0.059* | Nova 1956 | | 18.5# | 5200 | 3.9 | | 23, 24 |
| GQ Mus | 0.0594 | Nova 1983 | | 18.5# | 5000 | 4.0 | ** | 25, 26 |
| V1974 Cyg[§] | 0.0813 | Nova 1992 | | 16.2# | 1700 | 4.0 | ** | 27, 28, 29 |

References: (1) P11; (2) this paper; (3) Zellem et al. 2009; (4) Nogami et al. 2003; (5) Ohshima et al. 2012; (6) de Miguel et al. 2013; (7) Thorstensen et al. 1997; (8) Patterson et al. 1995; (9) Robertson et al. 1995; (10) Olech et al. 2008; (11) Fried et al. 1999; (12) Rutkowski et al. 2009; (13) Harvey et al. 1995; (14) Kato et al. 2002; (15) Pavlenko et al. 2010; (16) Olech et al. 2004; (17) Ishioka et al. 2001; (18) Kato et al. 2003; (19) Olech et al. 2007; (20) Urban & Sion 2006; (21) O'Donoghue et al. 1989; (22) Bianchini et al. 2012; (23) Retter & Lipkin 2001; (24) Tamburini et al. 2007; (25) Hachisu et al. 2008; (26) Diaz et al. 1995; (27) Hachisu & Kato 2005; (28) Chochol et al. 1997; (29) Skillman et al. 1997; (30) Patterson & Warner 1998; (31) Wood et al. 2011; (32) Kato 1993; (33) Olech et al. 2009; (34) Antonyuk & Pavlenko 2005; (35) Coyne et al. 2012; (36) Cannizzo et al. 2012; (37) Thorstensen & Taylor 1997; (38) Still et al. 2010; (39) Otulakowska-Hypka et al. 2012.

NOTES:
* $P_{orb}$ not precisely known (possible confusion with superhumps – or, less likely, some other non-orbital clock).
** Supersoft X-ray source (T>200000 K) for several years after eruption; $T_{wd}$ presently unknown, but likely very high.
*** We omit a few recent novae with less substantial data, or still somewhat in the throes of eruption: DD Cir (1999), V2362 Cyg (2006), and V458 Vul (2007).
# In 2012; star apparently still in slow decline from the eruption (SC).
[§] Possessing negative superhumps, with a morphology consistent with that of BK Lyn (i.e., primarily visible when faint). A few others, not flagged, have *possible* signals of this type.
[†] ER UMa distances are primarily from the dwarf-nova standard-candle relation (P11); but none of the calibrators are actually ER UMa stars, so some extra caution is warranted. Errors are probably 25-35%.

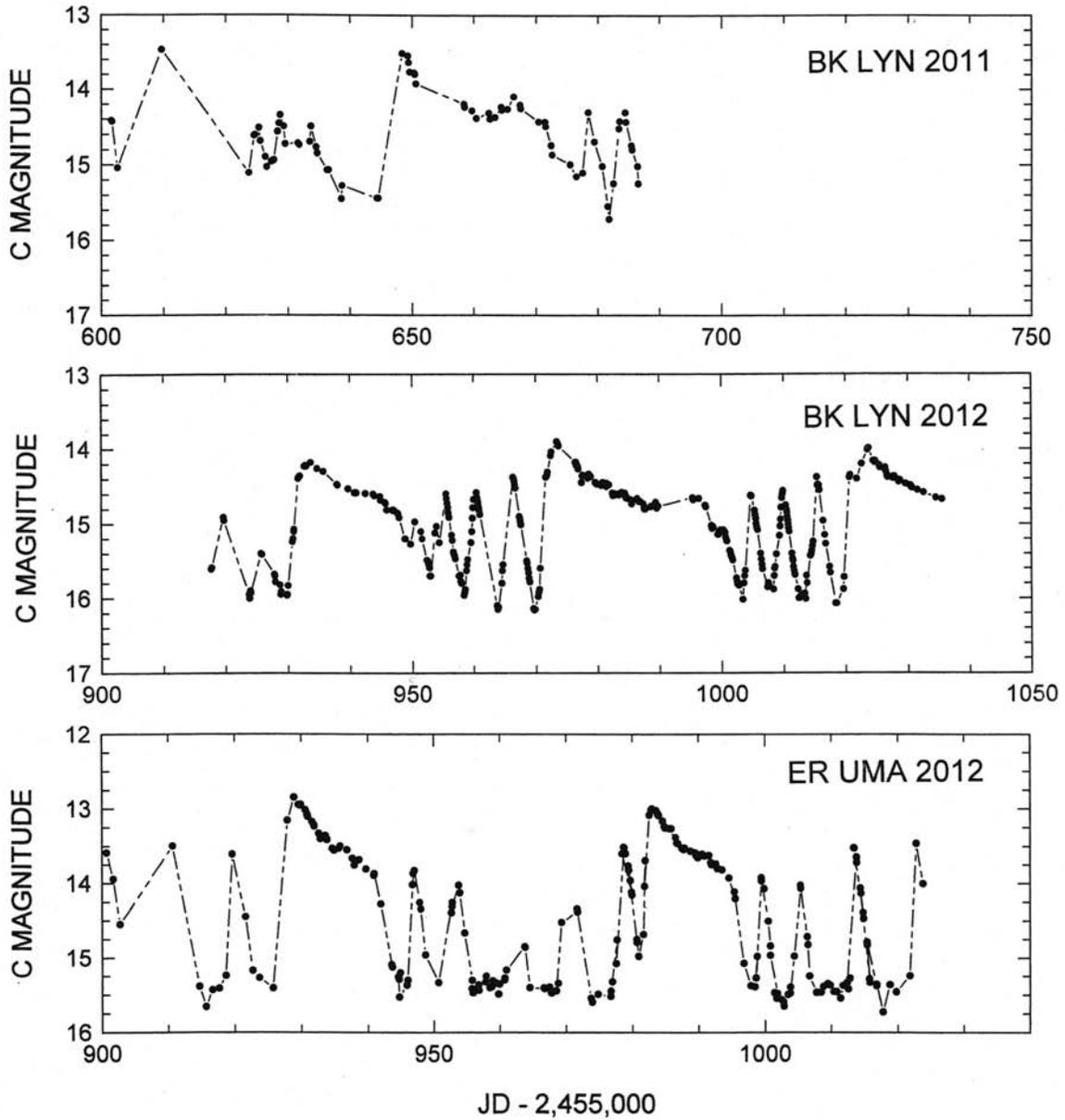

Figure 1. Top two frames: BK Lyn's seasonal light curve in 2011 and 2012. Bottom frame: 2012 seasonal light curve of the dwarf nova ER UMa. All points are averages over 1-3 orbital periods. (This is important, because snapshot magnitudes are polluted by the periodic signals and by erratic flickering.) In 2012, BK Lyn's light curve was indistinguishable, even in fine detail, from that of ER UMa, a prototype dwarf nova.

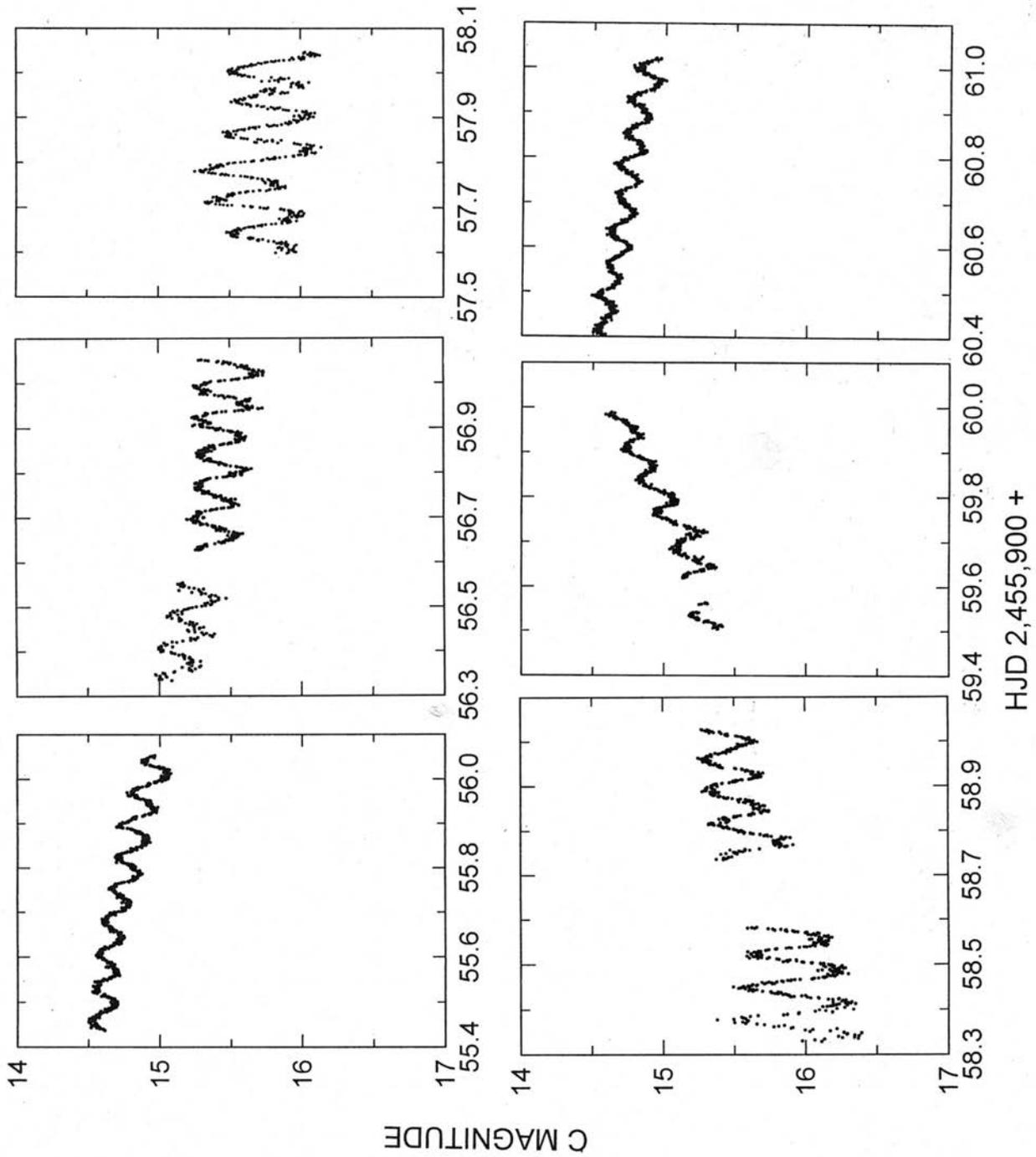

Figure 2. One 6-day segment of the 2012 light curve, illustrating the rapid variability during one cycle of "normal" outburst. The time scale is uniform across all frames. The obvious periodic signal is always most prominent near *minimum* light; in intensity units, the signal maintains a nearly constant amplitude.

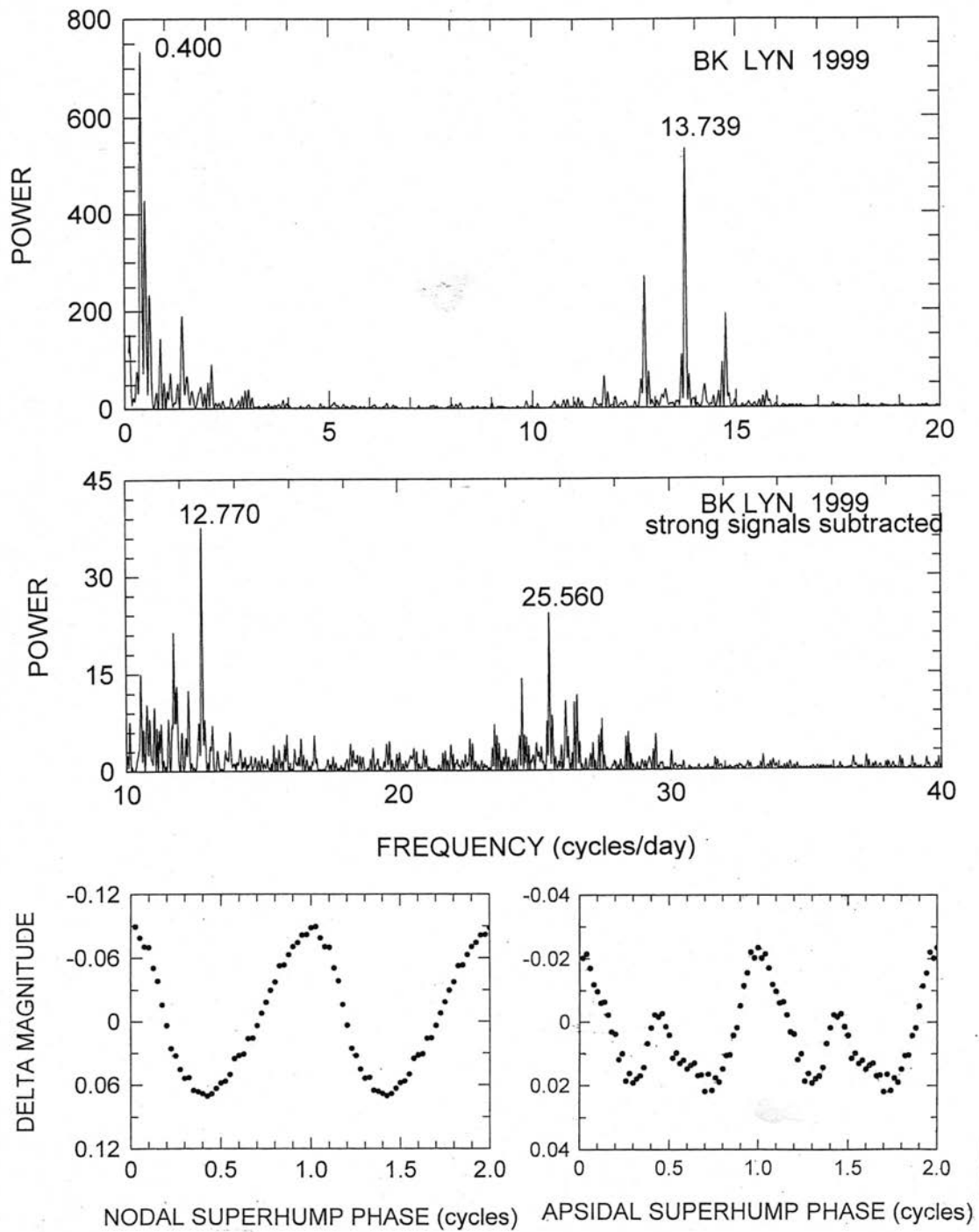

Figure 3. 1999 coverage.  *Upper frame*: the power spectrum of the 14-day light curve, showing two strong signals – marked by their frequency in cycles/day.  *Middle frame*: power spectrum of the residuals, after subtraction of the two strong signals; this shows a weaker signal at 12.77 c/d, with a significant second harmonic.  *Lower frame*: mean waveforms of the two superhump signals.

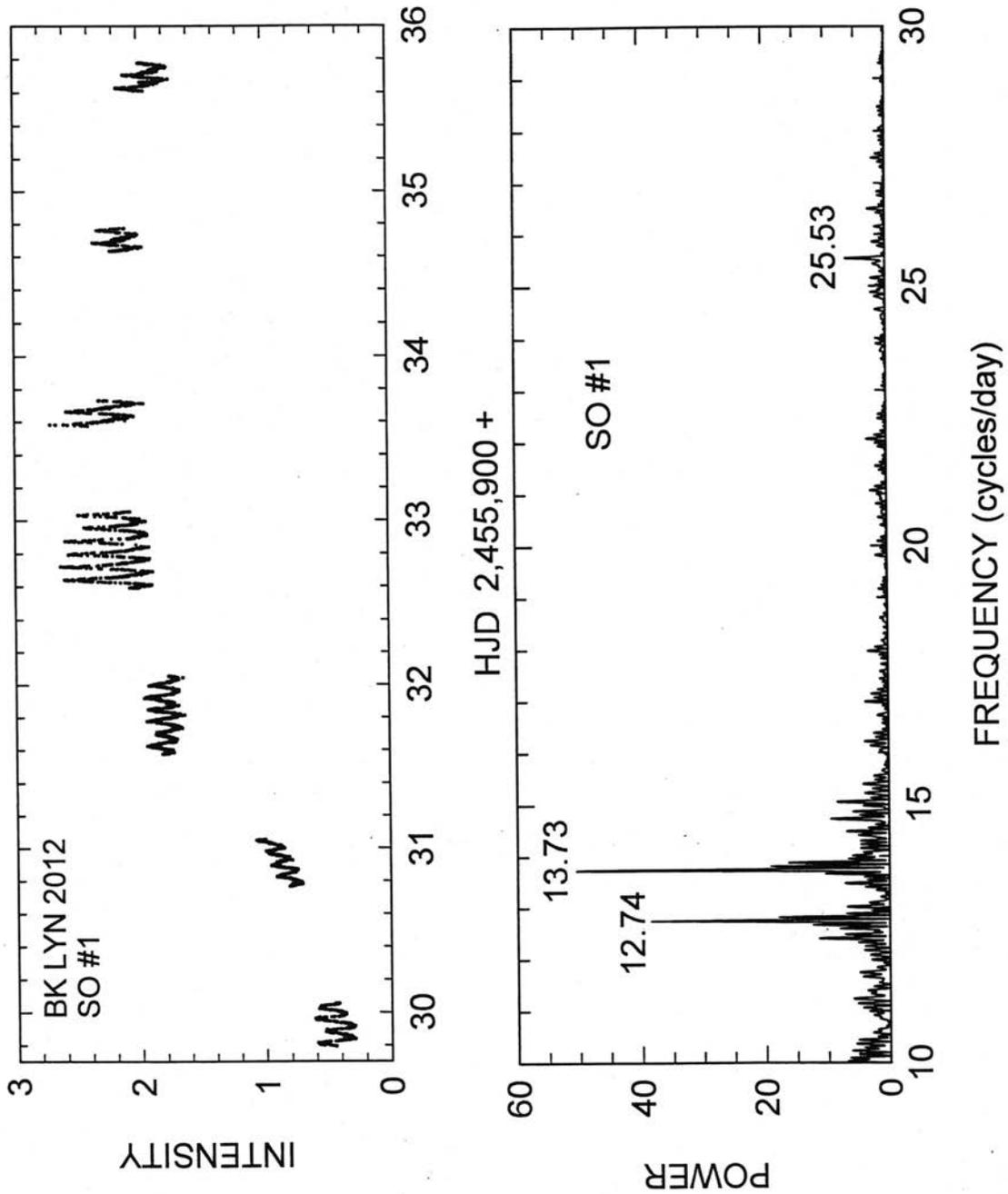

Figure 4. *Upper frame*: seven-day light curve near onset of the first superoutburst of 2012, in (arbitrary) intensity units. This shows the sudden growth of the apsidal superhump. *Lower frame*: power spectrum of that (11-day) superoutburst time series – showing the same pattern seen in 1999.

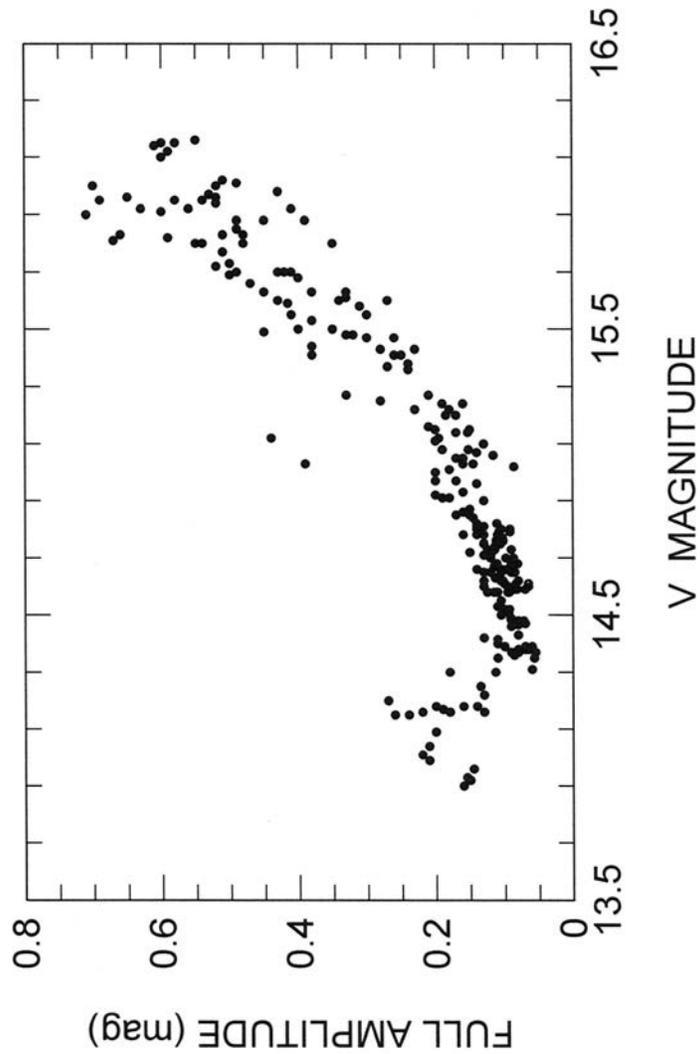

Figure 5. Full amplitude of the 2012 superhump signal and its dependence on *V* magnitude. This mostly arises from the negative superhump, which is most prominent when the star is faint – roughly constant, in intensity units, regardless of the outburst state. But near the peak of superoutburst (*V*≈14), the amplitude is much higher, due to confusion with the suddenly hatched positive superhump.

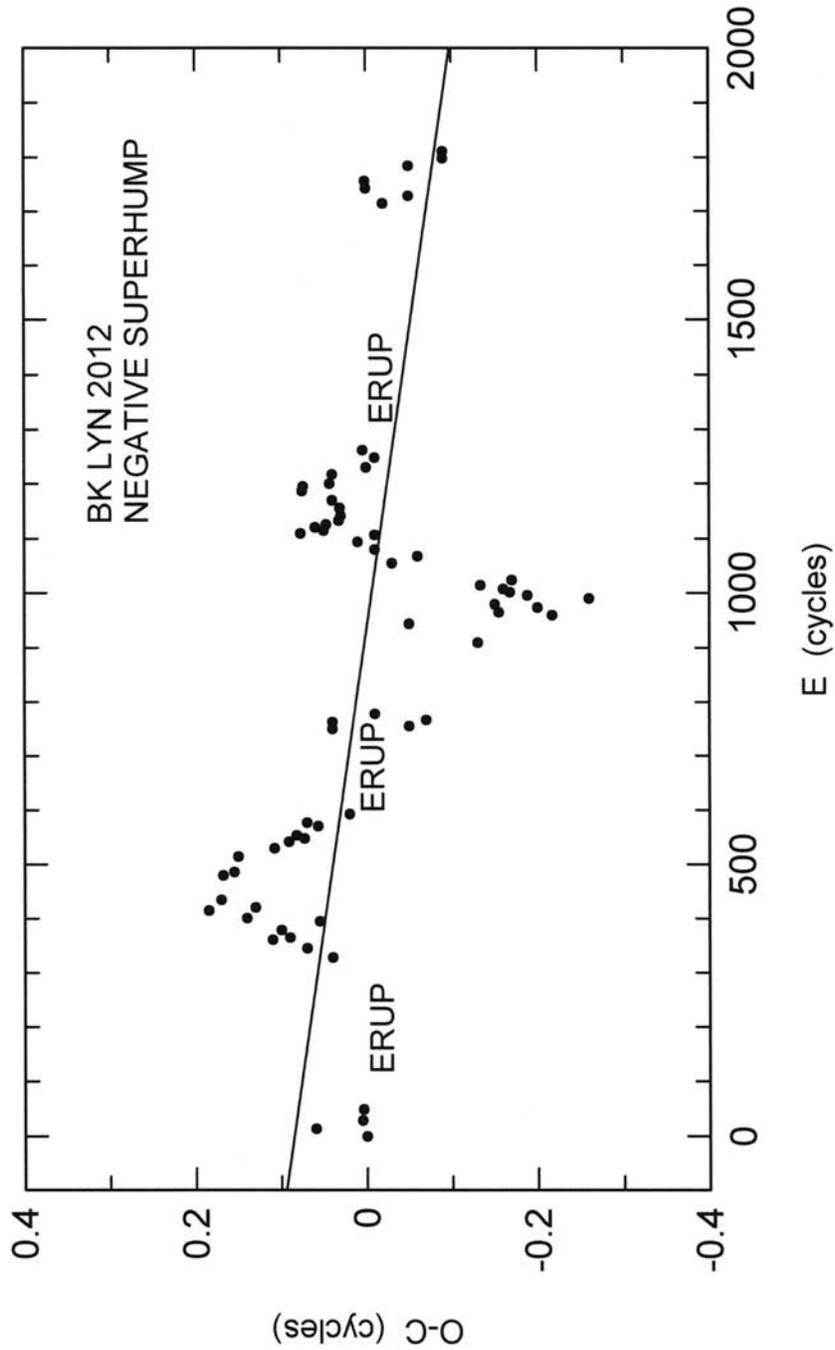

Figure 6. O-C diagram of the timings of (negative) superhump maxima, relative to the test ephemeris HJD 2,455,928.7875 + 0.072855E. Each timing represents one superhump cycle, and each displayed point is the average of 2-5 timings. Timings during eruption are significantly contaminated by the positive superhump, and are therefore omitted. The best linear fit is HJD 2,455,928.797 + 0.072846E, with some wanderings of ~0.1 cycle.

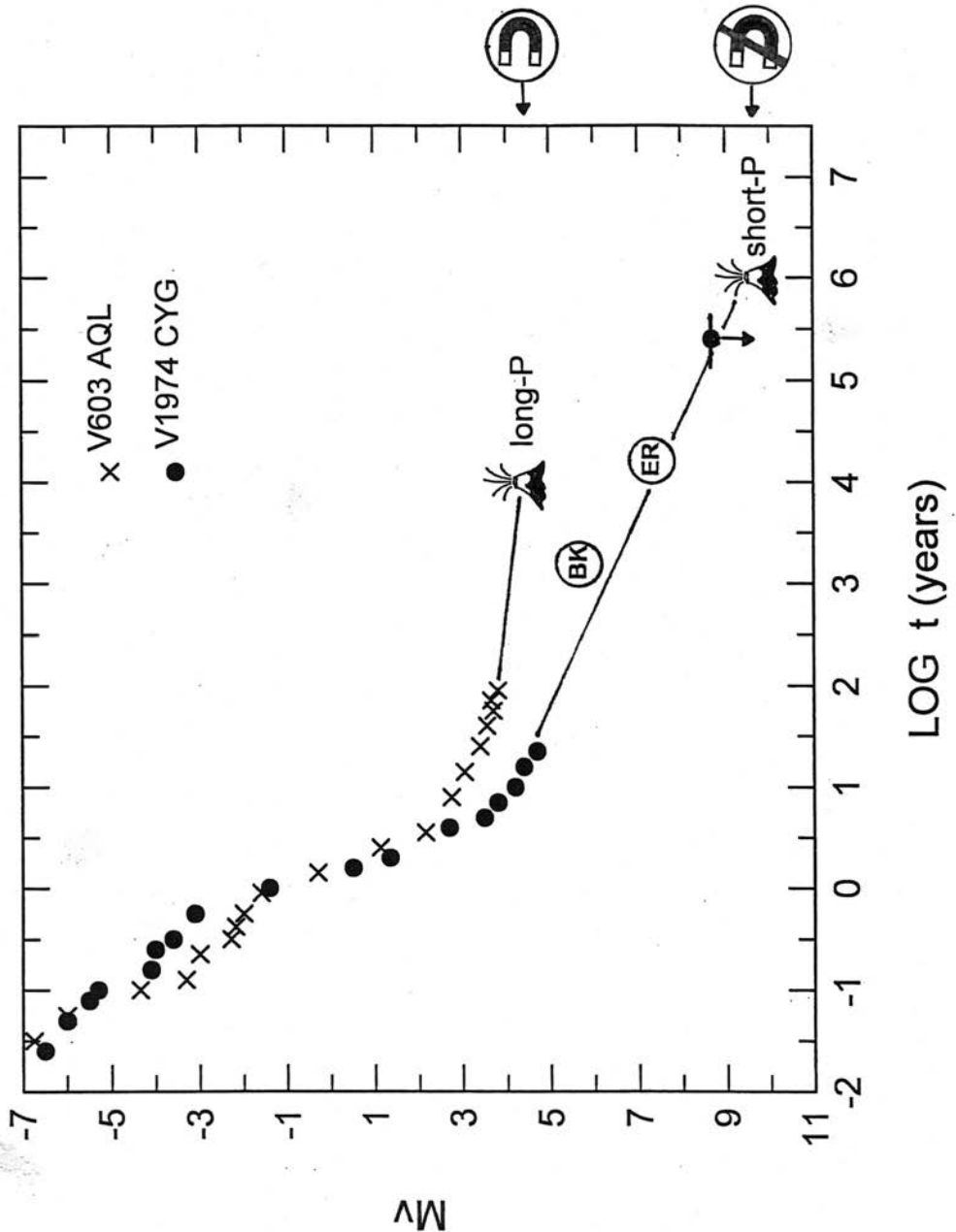

Figure 7. Roadmap for the decline of fast classical novae, for stars with $P_{orb}$ <2.4 hr ("short") and 2.8 hr <$P_{orb}$<10.0 hr ("long"). Dots and crosses show the visual magnitude history of two novae taken to be representative, V603 Aql and V1974 Cyg (including the latter's $M_v$>+9 limit from its pre-outburst nondetection). The two named dwarf novae are BK Lyn and the average of the other seven ER UMa stars in Table 4 (assuming the 2000 and 15000 yr ages, respectively, argued for in this paper). Long-$P_{orb}$ stars are assumed to be driven by magnetic braking, which will produce an eruption in ~$10^4$ years since it provides $\dot{M}$ ~$10^{-8}$ Mo/yr. Short-$P_{orb}$ stars must wait ~$10^6$ years, since they must rely on gravitational radiation, which provides only $10^{-10}$ Mo/yr.

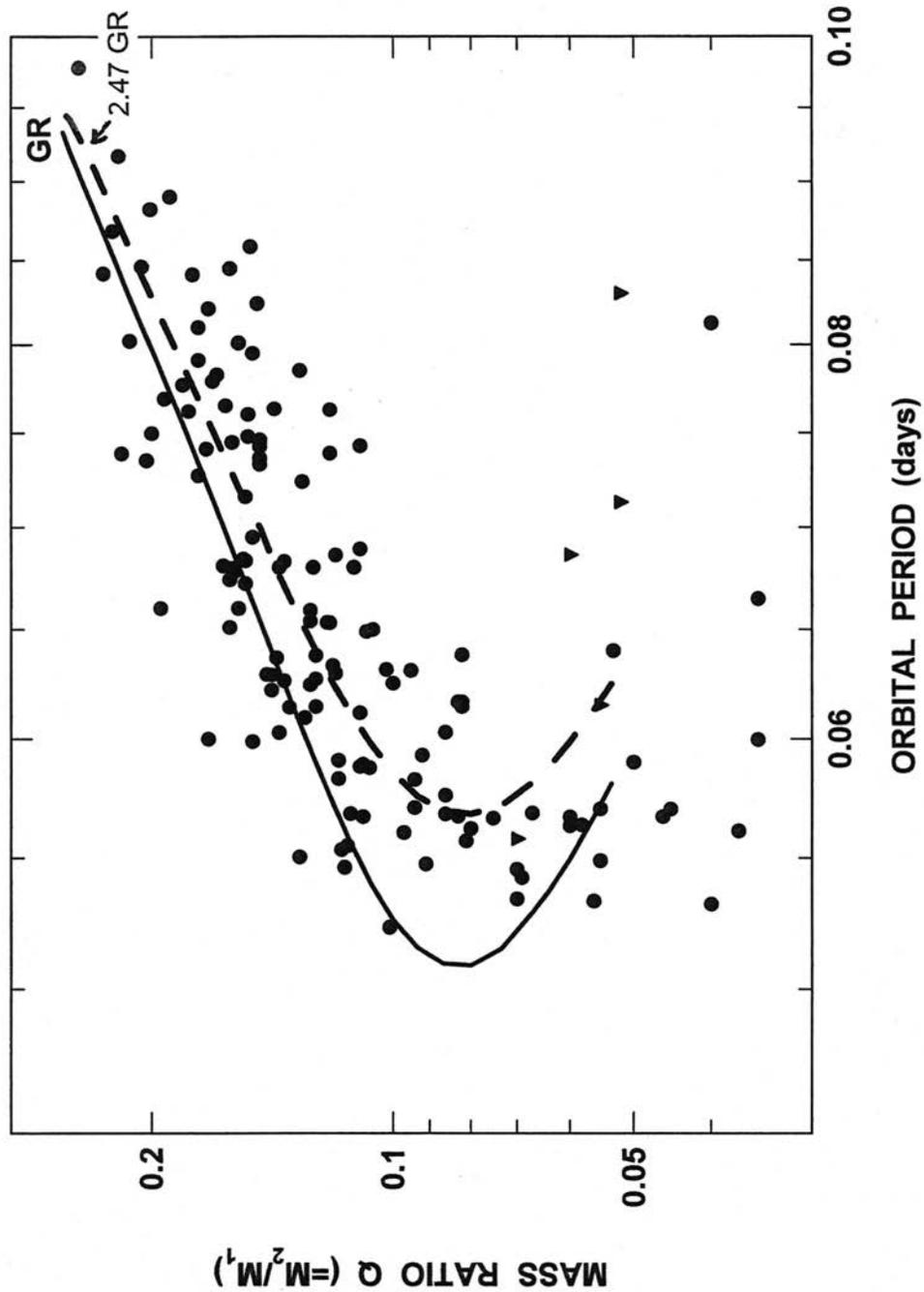

Figure 8. $q(P_{orb})$ correlation, compared to the predicted trend if evolution is driven purely by GR (solid curve). Data are taken from Table 2 of P11. Triangles are upper limits on $q$. The minimum $P_{orb}$ is ~10% longer than predicted, and the mean value of $P_{orb}$ at each $q$ is somewhat longer. These discrepancies are ameliorated if the angular-momentum loss is 2.47x that produced by GR alone (KBP, dashed curve).

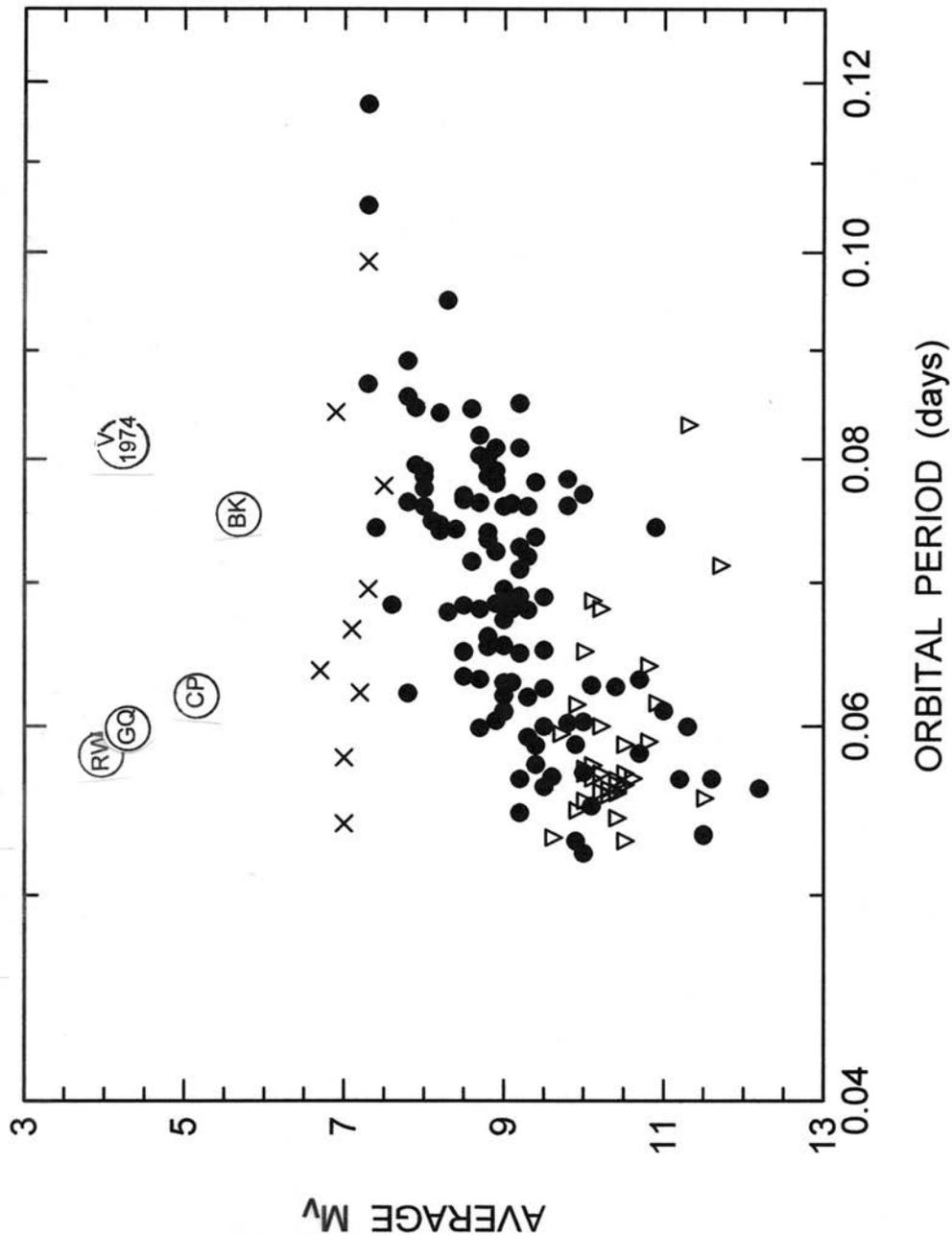

Figure 9. Time-averaged $M_v$ for short-$P_{orb}$ CVs, with data taken largely from Table 2 of P11. Nova remnants and BK Lyn are labelled by name, and the ER UMa stars are labeled with crosses. Triangles show upper limits on brightness, usually based on lower limits on a dwarf nova's recurrence time. The hypothesis of this paper is that each star suddenly jumps up to $M_v \approx -6$ in a classical-nova eruption, then falls vertically downward: to +4.5 in year 40, +6 in year 2000, +7 in year 15000, probably +8 in year 100000, and probably "quiescence" near +9 in year 300000. This roughly follows a $dM_v/d(\log t) = 1.0$ law. Except for lower-branch stars (period bouncers), the main source of dispersion at a given $P_{orb}$ is *time since the last nova eruption*.

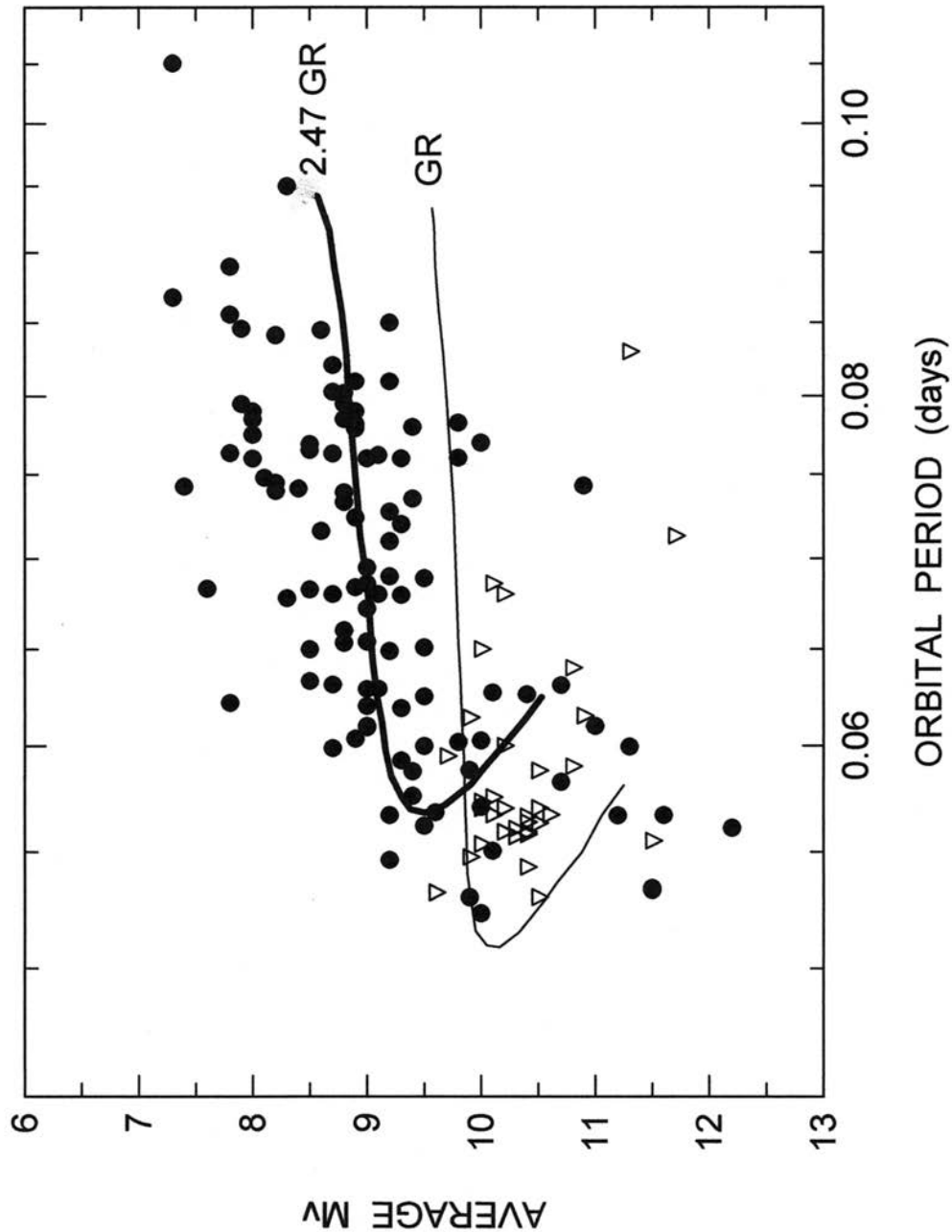

Figure 10. A closer look at $M_v(P_{orb})$ for the main cluster of stars in Figure 9, compared with predictions from evolution models (assuming a bolometric correction of 1.4 mag, appropriate for erupting dwarf novae). GR (thin curve) does a somewhat poor job at reproducing the observed trends. The bold curve is the KBP fit [to $R_2(P_{orb})$] of 2.47 GR. This does a much better job for $P_{orb}>0.06$ d, but appears to miss minimum $P_{orb}$ on the high side. (In golf this is called the "professional" side... so, you know, it has to be right.) Some enhancement to GR seems warranted, but its mathematical form and physical origin are still elusive.